\documentclass[12pt]{article}
\usepackage{amssymb,graphicx,amsmath,array,verbatim}

\makeatletter
\@addtoreset{equation}{section}

\renewcommand{\thefootnote}{\fnsymbol{footnote}}
\makeatother

\newcommand {\beq}{\begin{eqnarray}}
\newcommand {\eeq}{\end{eqnarray}}
\def\p{\partial}
\newcommand{\NF}{N_{\rm F}}
\newcommand{\NC}{N_{\rm C}}
\newcommand{\vs}[1]{\vspace{#1 mm}}
\newcommand{\hs}[1]{\hspace{#1 mm}}
\newcommand{\bpm}{\begin{pmatrix}}
\newcommand{\epm}{\end{pmatrix}}

\newcommand{\R}{\mathbb{R}}
\newcommand{\C}{\mathbb{C}}

\newcommand{\tr}{{\rm Tr}}
\newcommand{\D}{\mathcal D}

\newcommand{\ba}{\left( \begin{array}}
\newcommand{\ea}{\end{array} \right)}
\newcommand{\be}{\begin{equation}}
\newcommand{\ee}{\end{equation}}
\newcommand{\bea}{\begin{eqnarray}}
\newcommand{\eea}{\end{eqnarray}}
\newcommand{\beann}{\begin{eqnarray*}}
\newcommand{\eeann}{\end{eqnarray*}}
\newcommand{\nn}{\nonumber}

\newcommand{\Tr}{{\rm Tr}}

\newcommand{\del}{\partial}

\def\cprime{$'$} \def\cprime{$'$}

\begin{document}

\begin{titlepage}

\setcounter{page}{0}
\renewcommand{\thefootnote}{\fnsymbol{footnote}}

\begin{flushright}
TIT/HEP--581\\
TU--813\\
UT-08-06\\
May, 2008
\end{flushright}

\vspace{0mm}
\begin{center}
{\LARGE \bf Intersecting Solitons \\ Amoeba 
and Tropical Geometry\\
}

\vspace{15mm}
{\normalsize\bfseries
Toshiaki~Fujimori$^1$, 
Muneto~Nitta$^2$, \\ 
Kazutoshi~Ohta$^3$,
Norisuke~Sakai$^4$
and
Masahito~Yamazaki$^5$
}

\footnotetext{
e-mail~addresses: 
fujimori(at)th.phys.titech.ac.jp,
nitta(at)phys-h.keio.ac.jp, 
kohta(at)phys.tohoku.ac.jp,
sakai(at)lab.twcu.ac.jp,
yamazaki(at)hep-th.phys.s.u-tokyo.ac.jp
}

\vskip 1.5em

{ \it
$^1$Department of Physics, Tokyo Institute of 
Technology \\
Tokyo 152-8551, JAPAN  \\[2mm]
$^2$Department of Physics, Keio University, Hiyoshi,
Yokohama, Kanagawa 223-8521, JAPAN
  \\[2mm]
$^3$Department of Physics, 
Tohoku University, 
Sendai 980-8578, JAPAN 
\\[2mm]
$^4$Department of Mathematics, Tokyo Woman's Christian University, 
Tokyo 167-8585, JAPAN 
\\[2mm]
$^5$Department of Physics, University of Tokyo, Tokyo 113-0033, JAPAN 
}

\end{center}

\vspace{12mm}

\centerline{{\bf Abstract}}
We study generic intersection (or web) of vortices 
with instantons inside, 
which is a 1/4 BPS state in the Higgs phase of five-dimensional 
$\mathcal{N}=1$ supersymmetric $U(\NC)$ gauge theory on 
$\mathbb{R}_t \times (\C^{\ast})^2 \simeq {\mathbb R}^{2,1} \times T^2$ 
with $\NF=\NC$ Higgs scalars in the fundamental representation. 
In the case of the Abelian-Higgs model ($\NF=\NC=1$), 
the intersecting vortex sheets can be beautifully understood 
in a mathematical framework of amoeba and tropical geometry, 
and we propose a dictionary relating solitons and gauge theory 
to amoeba and tropical geometry. 
A projective shape of vortex sheets is described by 
the amoeba. Vortex charge density is uniformly 
distributed among vortex sheets, and negative contribution to 
instanton charge density 
is understood as the complex Monge-Amp\`ere measure 
with respect to a plurisubharmonic function on $(\C^{\ast})^2$.
The Wilson loops in $T^2$ are related with derivatives of 
the Ronkin function. 
The general form of the K\"ahler potential and the 
asymptotic metric of the moduli space of 
a vortex loop are obtained as a by-product. 
Our discussion works generally 
in non-Abelian gauge theories, 
which suggests 
a non-Abelian generalization of 
the amoeba and tropical geometry.

\vspace{10mm}

\end{titlepage}
\newpage

\renewcommand{\thefootnote}{\arabic{footnote}}
\setcounter{footnote}{0}

\section{Introduction}
The study of topological solitons is intimately connected 
with the development of mathematics. 
Originally solitons are found as solutions to non-linear 
(field) equations in various physical systems, but many 
mathematical concepts have been developed at the same time 
in order to understand properties and 
integrability of the soliton equations. 
Indeed, many mathematical tools, such as 
representation theory of differential operator algebra 
and infinite dimensional Grassmannian, 
are required to investigate the solitons. 
The language of two-dimensional conformal field theory 
is also useful in studies of a class of the soliton systems, 
such as KdV, Toda, KP equations. 
The investigation of solitons in gauge theory, 
such as instantons and monopoles, 
gives not only non-perturbative informations 
about the field theory, but also stimulates
new developments of differential geometry.

The purpose of the present article is to study topological 
solitons using novel mathematical objects known as amoeba 
and tropical geometry. 
We find a one-to-one correspondence between amoeba/tropical 
geometry and solitons in Yang-Mills-Higgs theory. 
Yang-Mills gauge fields coupled to the Higgs fields 
naturally appear as a bosonic part in supersymmetric 
Yang-Mills theory with eight supercharges.  
When sufficient number of the Higgs fields get vacuum 
expectation values (vevs), 
the theory is in the Higgs phase with completely broken gauge symmetry. 
Typical solitons in the Higgs phase known for a long time 
are vortices in the Abelian $U(1)$ gauge theory 
coupled to a single complex Higgs field 
(the Abelian-Higgs model) \cite{Abrikosov:1956sx}.
These vortices have been recently extended to the non-Abelian case, 
vortices in completely broken non-Abelian $U(N)$ gauge symmetry 
\cite{Hanany:2003hp}--\cite{Eto:2008yi}.
The other fundamental solitons found relatively recently in 
Yang-Mills-Higgs theory (or corresponding nonlinear sigma models)
are domain walls or kinks 
\cite{Eto:2006uw}, 
\cite{Abraham:1992vb}--\cite{Eto:2004vy}. 
Since both vortices and domain walls preserve half of the 
supercharges when embedded into supersymmetric theories, 
they are called 1/2 BPS solitons. 
Composite solitons in the Higgs phase have recently been studied 
extensively, especially in supersymmetric gauge theories 
with eight supercharges \cite{Tong:2005un,Eto:2006pg,Shifman:2007ce}. 
Since the magnetic field has to vanish in the Higgs phase,
magnetic monopoles are confined by vortices 
(confined monopoles) 
\cite{Tong:2003pz}--\cite{Eto:2004rz}.
Although isolated 
instantons shrink to points in the Higgs phase, 
they can lie inside a vortex core (trapped instantons) 
\cite{Hanany:2004ea,Eto:2004rz}.  
Vortex-strings can end on a domain wall \cite{Gauntlett:2000de}
or stretch between domain walls \cite{Isozumi:2004vg}. 
When a (composite) soliton configuration breaks 
$n$ directions of translational symmetry, it is defined to 
have $n$ codimensions. 
The composite soliton with the lowest co-dimension is 
a web or a network of domain walls, whose moduli space 
and dynamics have been 
worked out recently \cite{Eto:2005cp}-\cite{Eto:2007uc}. 
All of these composite solitons preserve 1/4 of 
supercharges if we realize the Yang-Mills-Higgs theory as 
a supersymmetric Yang-Mills theory, and are called 1/4 
BPS states. 
It has been found that all these 1/4 BPS composite 
solitons are related by the (Scherk-Schwarz twisted) 
dimensional reduction, starting from 
the instanton-vortex system \cite{Eto:2004rz}. 
Therefore understanding the instanton-vortex system 
is of primary importance to study 
BPS solitons in supersymmetric Yang-Mills-Higgs theories 
with eight supercharges.\footnote{
It is worth pointing out that there exist 
the other series of 1/4 BPS systems. 
This contains a triple intersection of vortices where 
a set of two vortices has one common codimension 
\cite{Naganuma:2001pu}, 
contrary to the instanton-vortex system \cite{Eto:2004rz} 
where vortices have no (or two) common codimensions. 
The former preserves (1,1) SUSY and the latter (2,0) SUSY 
in terms of two-dimensional supersymmetry \cite{Eto:2005sw}. 
Different sets of 1/4 BPS equations in these two series 
are obtained as dimensional reductions of the unique set of 
the 1/8 BPS equations \cite{Eto:2005sw,Lee:2005sv}. 
}
Unfortunately, generic configurations of instantons 
and vortex sheets as co-dimension four solitons have 
not been worked out, apart from a trapped instanton as a lump 
on a (uncurved) vortex plane, or an intersection point 
of two orthogonal (uncurved) vortex planes \cite{Eto:2004rz}. 
It is important to characterize generic configurations of 
instantons and vortex sheets as co-dimension four 
solitons in a precise and transparent manner. 
We can call these solitons as webs of vortices. 
We show that these soliton webs are nicely described 
in terms of the amoeba and the tropical geometry.

Interestingly, amoeba and tropical geometry 
have already appeared in physics literature, 
in the context of topological strings. 
Topological string amplitude (or its building block, topological vertex) 
can be described by means of a melting crystal picture 
\cite{Okounkov:2003sp,Iqbal:2003ds}, which was 
originally introduced as a statistical model in mathematical physics. 
Shape of the melting crystal in the thermodynamic 
limit is well interpreted in terms of an amoeba, which 
is a logarithmic projection of a smooth Riemann surface. 
On the other hand, an emaciated body of the amoeba 
corresponding to zero temperature limit can be captured 
by tropical geometry, where we can obtain some 
properties of the Riemann surface from a skeleton of the amoeba.
In the context of superstring theory, the amoeba and 
tropical geometry appear as, respectively, 
quantum ($g_s \gg 1$) and classical ($g_s \to 0$) shape of 
intersecting five-branes (five-brane junctions), 
which are dual to a suitable Calabi-Yau geometry where 
the topological string is defined. 
Other examples appear in an instanton and BPS state 
counting problem of the supersymmetric (quiver) gauge 
theories \cite{Nekrasov:2003rj,Feng:2005gw,Maeda:2006we}, since these 
systems are closely related to the topological 
string amplitudes and realized in terms of the five-brane web.

The five-brane system is useful to understand the relationship 
between the gauge and string theory, and the amoeba and 
tropical geometry, which is the main subject of this article. 
Our study is inspired by these successful applications 
of the amoeba and tropical geometry to physics. 
We find a similarity between the five-brane web and 
the 1/4 BPS composite solitons of vortex sheets and instantons. 
Some of properties of the webs of vortices are still unclear. 
In particular, the instanton charge appears at the 
intersection point of the vortices, but we have not obtained 
a tool to see a distribution of the extra instanton 
charge on the vortex web. 
The five-brane web in the superstring theory and vortex web 
possess common properties. Indeed, 
toric diagrams and geometry play an important role in 
both sides, and their moduli spaces are described by 
similar quotient spaces and the moduli space of the vortex 
web should be included in that of the five-brane web.
So we expect that the amoeba and tropical geometry, which 
is important to understand dynamics of the five-brane web, 
are also useful to analyze the web of the solitons in the 
Yang-Mills-Higgs system.

In this paper, we study 
the most generic configurations of 1/4 BPS solitons of 
instantons and vortices in the Higgs phase of the 
five-dimensional $\mathcal{N}=1$ supersymmetric $U(\NC)$ 
gauge theory (with eight supercharges) on 
$\mathbb{R}_t \times (\C^{\ast})^2 \sim \mathbb{R}^{2,1} \times T^2$ 
with $\NF=\NC$ Higgs scalars in the fundamental representation, 
by using the moduli matrix formalism \cite{Eto:2006pg}. 
Torus $T^2 = S^1\times S^1$ allows us to obtain other 1/4 BPS 
solitons readily through dimensional reduction. 
We show that vortex sheets are defined by zeros of 
a Laurent polynomial of two complex coordinates of $(\C^\ast)^2$, 
and instanton positions are given by common zeros with 
another polynomial. 
We find that the above expectation of the importance of 
amoeba and tropical geometry to be correct. 
We describe physical quantities of the intersecting 
solitons (soliton web) 
in terms of the mathematical language of amoeba and tropical geometry.
We also see the important objects in the 
amoeba and tropical geometry, such as the logarithmic 
mapping, the Ronkin function and the Monge-Amp\`ere measure, also 
have essential meanings on the soliton side. 
We describe properties of the soliton web in terms of these 
mathematical objects.
We find that the moduli matrix approach is very useful in 
the translation between physical and mathematical languages. 

The organization of the present paper is as 
follows. 
In Sec.\,\ref{sec:higgs_phase}, we review the 1/4 BPS equations for 
vortices and instantons in the Higgs phase of 
five-dimensional $\mathcal{N}=1$ supersymmetric $U(\NC)$ 
gauge theory on 
$\mathbb{R}_t \times (\C^{\ast})^2 \sim \mathbb{R}^{2,1}\times T^2$ 
with $\NF=\NC$ Higgs scalars in the fundamental representation, 
using the moduli matrix formalism. 
We mainly consider vortices in the overall $U(1)$ gauge theory 
except in Sec.\,\ref{sec:instanton}. 
In Sec.\,\ref{sec:cylinder}, we consider simpler case of 
vortex on cylinder 
$\R \times S^1 \simeq \C^\ast$. 
In Sec.\,\ref{sec:vortex_sheets_C2}, we study the most general situation of 
vortex sheets on $(\C^\ast)^2$. 
Sec.\,\ref{sec:amoeba} gives amoeba corresponding to vortex sheets. 
Sec.\,\ref{sec:tropical_geomegtry} relates it to the tropical 
geometry and gives an example. 
Sec.\,\ref{sec:topological_charge} describes a general formula 
to compute the topological charges. 
Sec.\,\ref{sec:metric} gives the metric of the moduli space 
of a vortex loop. 
We also obtained a new general formula for the K\"ahler metric 
which is valid for arbitrary values of $N_{\rm C}$ and $N_{F}$. 
In Sec.\,\ref{sec:instanton}, 
we examine the instanton number in non-Abelian gauge theory. 
Sec.\,\ref{sec:i-number1} reviews  
the instanton number 
for instantons trapped inside 
a non-Abelian vortex plane \cite{Eto:2004rz}.  
Sec.\,\ref{sec:i-number2} discusses more general configurations 
of the instantons trapped inside a non-Abelian vortex web.
Sec.\,\ref{sec:discussion} is devoted to conclusion and discussion.

\section{Vortices and Instantons}
\label{sec:higgs_phase}

In this section, we first review the construction 
of vortex solutions in $U(\NC)$ gauge theory 
in (4+1)-dimensional spacetime 
$\mathbb{R}_t \times (\C^{\ast})^2 \sim \mathbb{R}^{2,1}\times T^2$ 
with $\NF$ Higgs fields in 
the fundamental representation. 
By 
introducing additional 
$N_{\rm F}$ 
Higgs fields in the fundamental representation,
this theory can also be regarded as the bosonic part of a 
five-dimensional ${\cal N}=1$ supersymmetric $U(\NC)$ 
gauge theory with $\NF$ hypermultiplets in the fundamental 
representation, but the fermionic part 
(and another set of $N_{\rm F}$ Higgs scalars) 
is irrelevant and is omitted 
in the following discussion.
We introduce $(x_1,\,y_1,\,x_2,\,y_2)$ 
and $z_1 \equiv x_1 + i y_1 ,~z_2 \equiv x_2 + i y_2$ 
as real and complex coordinates of $(\C^\ast)^2$, respectively. 

The Lagrangian of the theory takes the form
\beq
\mathcal L = \tr \left[ - \frac{1}{2g^2} F_{\mu \nu} F^{\mu \nu} + \D_\mu H (\D^\mu H)^\dagger - \frac{g^2}{4} (HH^\dagger - c \mathbf 1_{\NC})^2 \right],
\label{eq:Lagrangian}
\eeq
where the Higgs fields are expressed as 
an $\NC \times \NF$ matrix $H^{rA}~(r=1,\cdots,\NC,~A=1,\cdots,\NF)$. 
The covariant derivative is defined by 
$\D_\mu H = \p_\mu H + i W_{\mu} H$ 
and the field strength by 
$F_{\mu \nu}=-i[\D_\mu,\D_\nu]=\p_\mu W_\nu - \p_\nu W_\mu + i [W_\mu,W_\nu]$. 
The constants $g$ and $c$ are the gauge coupling constant 
and the Fayet-Iliopoulos (FI) parameter, respectively.
At the vacua (the minima of the potential) of this theory, 
the Higgs fields $H$ get vev 
and $U(N)$ gauge symmetry is completely broken. 
Namely the theory has only the Higgs branch due to 
the nonzero FI term. 
The moduli space of the vacua is given by a complex Grassmannian
\beq
G(\NF,\NC) = \frac{SU(\NF)}{SU(\NC) \times SU(\NF-\NC) \times U(1)}.
\label{eq:vacua}
\eeq

Considering a static gauge configuration, 
we find that there is a lower bound of the energy 
\cite{Hanany:2004ea,Eto:2004rz} 
\beq
E ~\geq~ - \frac{1}{g^2} \int \tr \left( F \wedge F \right) - c \int \tr \, F \wedge \omega
=\frac{8\pi^2}{g^2}I+2\pi c\, V,
\label{eq:BPS-bound}
\eeq
where the two form $\omega \equiv \frac{i}{2} 
(dz_1 \wedge d\bar z_1 + dz_2 \wedge d \bar z_2)$ is the 
K\"ahler form on $(\C^{\ast})^2$, and we have defined the 
total instanton charge $I$ 
as an integral of the instanton charge density $\mathcal{I}$, 
and the vortex charge $V$ as a divergent integral of the 
vortex charge density $\mathcal{V}$ 
\beq
 I &\equiv& \int \mathcal{I} ~\,\equiv~ 
- \frac{1}{8\pi^2} \int \tr \left( F \wedge F \right) 
 ~=~ 
 \int ch_2,  \label{eq:total-instanton} \\
 V &\equiv& \int \mathcal{V} ~\,\equiv~ 
- \frac{1}{2\pi} \int \tr \, F \wedge \omega 
 ~~~~~=~ \int c_1 \wedge \omega. \label{eq:vortex-charge} 
\eeq
The lower bound Eq.\,(\ref{eq:BPS-bound}) is saturated if 
the following BPS equations \cite{Hanany:2004ea,Eto:2004rz}
\beq
F_{\bar z_1 \bar z_2} = 0, \hs{10} 
\D_{\bar z_i} H = 0, \hs{10} 
-2i (F_{z_1 \bar z_1} + F_{z_2 \bar z_2}) 
= \frac{g^2}{2}(HH^\dagger - c \mathbf 1_{N_{\rm C}}),
\label{eq:BPS-equations}
\eeq
are satisfied.
When FI parameter $c$ is sent to zero, 
the Higgs field $H$ vanishes and these equations reduce to 
the anti-self-dual equations 
for Yang-Mills instantons, 
whereas when we neglect the $z_2$-(or $z_1$-)dependence of the fields 
they reduce to simple vortex equations for 
vortices on the $z_1$-(or $z_2$-)plane.  
These vortices are two-codimensional surfaces 
in the four dimensional space $(\C^{\ast})^2$.
Therefore the BPS equations 
(\ref{eq:BPS-equations}) contain at least 
instantons and intersecting vortex sheets. 
As we will see below, 
these equations describe
webs of vortex sheets in general. 
%
The equations (\ref{eq:BPS-equations}) 
were earlier found for those on 
arbitrary K\"ahler manifold \cite{MundetiRiera:1999fd} 
and were simply called ``vortex equations" although 
they contain instantons also.\footnote{
In Ref.~\cite{MundetiRiera:1999fd}
the vortex equations are defined on  
arbitrary K\"ahler manifold $M$ of 
complex dimension $n$ (with $n=2$ not necessary). 
There the bound is given by 
$\int_M \tr \left(F \wedge F \right)\wedge \omega^{n-2}$ and 
$\int_M \tr \, F \wedge \omega^{n-1}$ 
with the K\"ahler 2-form $\omega$,
instead of the charges (\ref{eq:total-instanton}) and 
(\ref{eq:vortex-charge}). 
Furthermore at least in the case of $N_{\rm C} = N_{\rm F}=1$ 
these generalized equations can be obtained as 
equivariant dimensional reduction of  
the Donaldson-Uhlenbeck-Yau equations on 
$M \times S^2$ with a monopole configuration on $S^2$ 
\cite{Popov:2005ik}. }
It has been shown in \cite{Eto:2004rz} that 
solutions to the BPS equations (\ref{eq:BPS-equations}) 
on $(\C^{\ast})^2$ (or $\C^2$) preserve a quarter of 
supercharges in the supersymmetric gauge theory 
with eight supercharges. 
So the configuration of the solution is called 
a 1/4 BPS state in this sense.
The energy of the BPS configuration is determined by the 
topological charges (\ref{eq:total-instanton}) and 
(\ref{eq:vortex-charge}).

The vortex charge $V$ can be evaluated from the 
sum of the area of each vortex sheet, as we will see 
in Sec.\,\ref{sec:vortex_sheets_C2}. 
The total instanton charge $I$ can be decomposed 
into the intersection charge $I_{\rm intersection}$ 
and the instanton number $I_{\rm instanton}$ as
\beq 
 && I = -  I_{\rm intersection} + I_{\rm instanton}, \phantom{\bigg[}\\
 && I_{\rm intersection} ~\equiv~ \int {\cal I}_{\rm intersection}
      ~\equiv~ \frac{1}{8\pi^2} \int \tr \, 
    F \wedge \tr \, F ~=~ \frac{1}{2} \int c_1 \wedge c_1,
      \label{eq:intersection}\\
 && I_{\rm instanton} ~~~\equiv~ \int {\cal I}_{\rm instanton}
       ~~~\equiv~ \int c_2.
      \label{eq:instantonnum}
\eeq 
The intersection charge has negative contribution 
to the energy of the BPS configuration, which  
can be regarded as binding energy of 
intersecting vortex sheets. 
On the other hand 
the instanton number (not to be confused with total 
instanton charge) 
counts to the number of usual (particle-like) instantons and 
has positive contribution to the energy. 

Let us solve the BPS equations (\ref{eq:BPS-equations}).
The first BPS equation $F_{\bar z_1 \bar z_2} \equiv 
- i [\D_{\bar z_1}, \D_{\bar z_2}] = 0$ in (\ref{eq:BPS-equations}) 
is equivalent to an integrability  
condition\footnote{
If we identity the Higgs field $H^{rA}$ as a set of $\NF$ 
sections of a rank $\NC$ vector bundle $E$ on the base 
K\"ahler manifold, then the first BPS equation 
$F_{\bar z_1 \bar z_2} = 0$ is equivalent to the condition 
of the existence of the holomorphic frame 
$\{\tilde e_i\}~(i=1,\cdots,\NC)$, which satisfy 
$\D_{\bar z} \tilde e_i =0$.
}
for the differential operators $\D_{\bar z_i}$, 
which states that there exists an $\NC \times \NC$ 
matrix-valued function\footnote{
The complexified gauge transformation ${S^i}_r$ can be 
interpreted as the change of basis from the unitary frame 
(orthonormal frame) $\{e_r\}~(r=1,\cdots,\NC)$ to the 
holomorphic frame $\{\tilde e_i\}$.
}  
$S(z_i,\bar{z}_i) \in U(\NC)^{\C} = GL(\NC,\C)$ 
such that 
\beq
W_{\bar z_i} = - i S^{-1} \p_{\bar z_i} S.
\label{eq:gauge-S}
\eeq
Defining an $\NC \times \NF$ matrix 
\beq 
H_0 \equiv S H,
\eeq 
the second equation in (\ref{eq:BPS-equations}) reduces to 
\beq
\p_{\bar z_i} H_0 = 0.
\label{eq:H_0}
\eeq
This means that the elements of 
the matrix $H_0$ should be 
holomorphic\footnote{
In other words, $H_0^{iA}$ is a set of $\NF$ holomorphic 
sections of the holomorphic vector bundle $E$.
} 
with respect to the complex coordinates $z_i$. 
The matrix-valued 
quantity $S$ is determined from the 
last equation in Eq.~(\ref{eq:BPS-equations}), 
which can be rewritten in terms of an $\NC \times \NC$ 
positive definite Hermitian matrix\footnote{
This matrix $\Omega$ can be interpreted as the inverse of 
the Hermitian metric in terms of the holomorphic frame 
$\{\tilde e_i\}$ which is the identity matrix in terms of 
the unitary frame $\{e_r\}$.
}
\beq 
\Omega \equiv SS^\dagger
\eeq 
into
\beq
\p_{\bar z_1} (\Omega \p_{z_1} \Omega^{-1}) 
+ \p_{\bar z_2} (\Omega \p_{z_2} \Omega^{-1}) 
= - \frac{g^2c}{4} \left( \mathbf 1_{\NC} - \Omega_0 \Omega^{-1} \right),
\label{eq:master}
\eeq
where we have defined 
\beq 
\Omega_0 \equiv \frac{1}{c} H_0 H_0^\dagger.
\eeq
We call the equation (\ref{eq:master}) the ``master equation" 
of the instanton-vortex system.\footnote{
When the FI-parameter $c$ goes to zero, 
the Higgs phase no longer exists. 
In this case, 
the RHS of Eq.\,(\ref{eq:master}) vanishes, 
and Eq.\,(\ref{eq:master}) becomes 
the so-called Yang's equation \cite{Yang:1977zf}
for usual instantons not accompanied 
by vortices. 
}

Using these redefined fields, 
we can solve the BPS equations 
by the following procedure. 
Take an arbitrary holomorphic matrix $H_0(z)$ 
and solve Eq.\,(\ref{eq:master}) in terms of $\Omega$,  
then we can determine $S$ up to $U(\NC)$ gauge transformation 
$S \rightarrow S U^\dagger$ and 
physical fields can be obtained via the relations
\beq
W_{\bar z_i} = -i S^{-1} \p_{\bar z_i} S, \hs{10} H = S^{-1} H_0.
\label{eq:solution}
\eeq
The equations (\ref{eq:H_0}) and (\ref{eq:master}) 
have a ``gauge symmetry'', which we call 
``$V$-transformation", defined by
\beq
(H_0,\,S) ~\rightarrow~ \left( VH_0,\,VS \right), \hs{10} V(z) \in GL(\NC,\C).
\label{eq:V-transformation}
\eeq
Note that the physical fields $W_{\bar z}$ and $H$ 
are invariant under the $V$-transformation, 
so this defines an equivalence relation called 
the ``$V$-equivalence".\footnote{
This equivalence relation originates from 
the redundancy of the holomorphic frame $\{\tilde e_i\}$. 
}
Assuming that there exists a unique solution of 
Eq.\,(\ref{eq:master}) for a given $H_0(z)$,\footnote{
This assumption is correct at least 
when the base space is compact and K\"ahler, 
since the uniqueness and existence of solutions to 
the BPS equations (\ref{eq:BPS-equations}) 
were rigorously proved in terms of 
the Hitchin-Kobayashi correspondence 
\cite{MundetiRiera:1999fd}.
}
we find that there exists a one-to-one correspondence 
between the equivalence class $H_0 \sim VH_0$ and 
a point on the moduli space of the BPS configurations. 
In this sense, we call $H_0(z)$ a ``moduli matrix" 
and the parameters contained in $H_0$ are 
identified with the moduli parameters of the BPS configurations. 

Now let us consider the case $\NC=\NF=N$ 
which is often called a local theory. 
If the determinant of the Higgs fields $H$ 
vanishes in some regions, 
the broken gauge symmetry is partially 
restored in those regions. 
As dictated by the Meissner effect in the Higgs phase, 
this gauge symmetry restoration occurs where the magnetic 
flux penetrates and a vorticity arises around the zero of 
the Higgs field. 
Therefore the vanishing determinant $\det H = 0$ defines 
a two-dimensional surface 
of vortex positions in the four-dimensional space: 
it is equivalently given by\footnote{
The determinant $\det H_0$ can be regarded as the 
holomorphic section of the determinant line bundle 
$\wedge^{\NC} E$, and the vortex sheet corresponds to 
the effective divisor associated with the holomorphic 
section $\det H_0$.
}
\beq
\det H_0(z_1,z_2) = 0.
\label{eq:vortex-position}
\eeq

In closing this section, 
we comment on the vortex solutions of 
the (2+1)-dimensional $U(\NC)$ gauge theory 
on $\mathbb{R}_t \times \C^\ast$ 
with $\NF$ massless Higgs fields in the fundamental representation, 
since we study this case in the next section. 
Historically, the non-Abelian vortices were 
first found \cite{Hanany:2003hp,Auzzi:2003fs} on $\C$  
in the color-flavor locked phase of $U(N_{\rm C})$ gauge theory
with $\NF$ massless Higgs fields in the fundamental representation.
After their discovery the non-Abelian vortices 
have been extensively studied by many authors 
\cite{NA-vortices}. In particular, the non-Abelian vortices 
on a cylinder $\C^\ast$ 
have been studied in \cite{Eto:2006mz,Eto:2007aw}.
In the moduli matrix formalism, the discussion 
of the $\C^\ast$ case is completely parallel to the 
$(\C^\ast)^2$ case 
and the necessary formulae are simply obtained 
by neglecting the $z_2, \bar{z_2}$ dependence. 
For example, the BPS bound \eqref{eq:BPS-bound} is reduced to 
\beq
 E\ge -c \int d^2 x\, \tr \, F_{xy}, \label{eq:BPS-bound-2}
\eeq
and the master equation \eqref{eq:master} for $\Omega(z_1)$ 
is given by \cite{Eto:2005yh}
\beq
 \p_{\bar z_1} (\Omega \p_{z_1} \Omega^{-1}) = - \frac{g^2c}{4} \left( \mathbf 1_{\NC} - \Omega_0 \Omega^{-1} \right).
 \label{eq:master-2}
\eeq
For the Abelian-Higgs model, $\NC=\NF=1$, 
this equation reduces to the so-called Taubes's equation \cite{Taubes:1979tm} 
after some redefinition.

\section{Vortices on a Cylinder $\C^\ast$}\label{sec:cylinder}

In this section, we first consider the simpler case of 
vortices on a cylinder $\R \times S^1 \simeq \C^\ast$ 
before discussing the intersecting vortices in four dimensions. 
The vortices are BPS solutions of 
the (2+1)-dimensional $U(\NC)$ gauge theory 
with $\NF$ massless Higgs fields in the fundamental representation, 
with one spatial direction compactified \cite{Eto:2006mz,Eto:2007aw}.
In the supersymmetric system, 
the vortices preserve a half of the supercharges.
Although our real interest is in complex two-dimensional case 
($\R^2\times T^2 \simeq (\C^\ast)^2$), 
one-dimensional case is simpler and useful 
to understand the discussion in the next section. 

Let $(x,y)$ and $z \equiv x + i y$ be real and 
complex coordinates of $\R \times S^1 \simeq \C^\ast$, respectively. 
The coordinate of $S^1$ has a period $2\pi R$, 
namely $y \sim y + 2 \pi R$.
We here concentrate on the case of $\NC=\NF=N$. 
The moduli matrix formalism works 
as well in this (2+1)-dimensional case, 
and the BPS solutions are parametrized 
by the moduli matrix $H_0(z)$.
Since the moduli matrix should satisfy 
the periodic boundary condition $H_0(z + 2 \pi i R) = H_0(z)$, 
the determinant of the moduli matrix 
can be expanded as a Fourier series
\beq
\det H_0 ~=~ \sum_{n \in \mathbb Z} a_n e^{n z/R}.
\label{eq:1dim-Fourier}
\eeq
Introducing a new coordinate $u \equiv e^{z/R}$, 
this can be rewritten as
\beq
\det H_0 ~=~ P(u) ~\equiv~ \sum_{n \in \mathbb Z} a_n u^n.
\label{eq:1dim-LP}
\eeq
The positions of the vortices are determined 
by zeros of this Laurent polynomial. 
By performing an appropriate $V$-transformation 
$H_0(z) \rightarrow V(z) H_0(z)$, 
$\det H_0$ reduces to
\beq
\det H_0 ~=~ \prod_{i=1}^k (e^{z/R} - e^{z_i/R}) ~=~ \prod_{i=1}^k(u-u_i), \hs{10} u_i \equiv e^{z_i/R}, 
\label{eq:Factor[det[H_0]]}
\eeq
with $k$ denoting the number of the vortices.
The solution $\Omega$ of the master equation (\ref{eq:master-2}) 
for the vortices approaches to $\Omega_0= \frac{1}{c} H_0H_0^\dagger$ 
in the strong gauge coupling limit $g \rightarrow \infty$. 
In this limit the configuration of the magnetic flux of the overall $U(1)$ 
becomes singular such as 
\beq
\tr \, F = i \bar \p \p \log \det \Omega ~\rightarrow~ 
i \bar \p \p \log |\det H_0|^2 
= - 2 \pi \sum_{i=1}^k \delta^2 (z-z_i) dx \wedge dy, 
\label{eq:g-inf-flux}
\eeq
with $\partial=d z \, \partial_z,~\bar \partial=d\bar z \, \partial_{\bar z}$. 
This reflects the fact that the size of the vortices 
is proportional to $l \equiv 1/g \sqrt{c}$ 
and becomes zero in the strong coupling limit\footnote{The appearance of length scale $l \equiv 1/g \sqrt{c}$ is understood from the master equation \eqref{eq:master-2}. In \eqref{eq:master-2}, the parameters $g$ and $c$ appear only in the combined form $g^2c$, and thus $l$ is the only length scale made from $g^2 c$.}. Using the configuration in the infinite coupling limit, 
we find that the topological charge is given by the number of zeros 
of the polynomial $P(u)$ 
\beq
- \frac{1}{2\pi} \int d^2 x \, \tr \, F_{xy} = k.
\label{eq:vortex-number}
\eeq

\begin{figure}[htbp]
\begin{center}
\begin{tabular}{ccc}
\includegraphics[width=50mm]{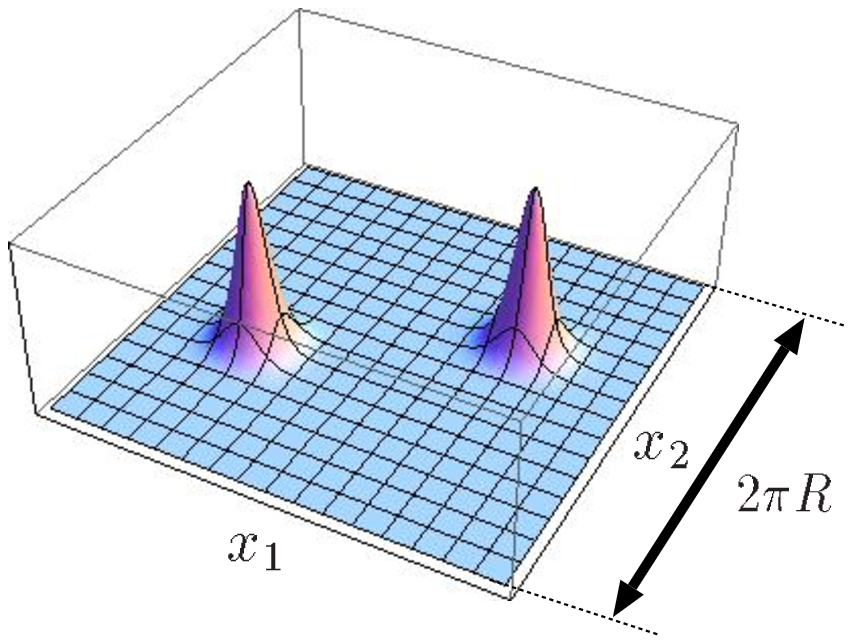} &
\includegraphics[width=50mm]{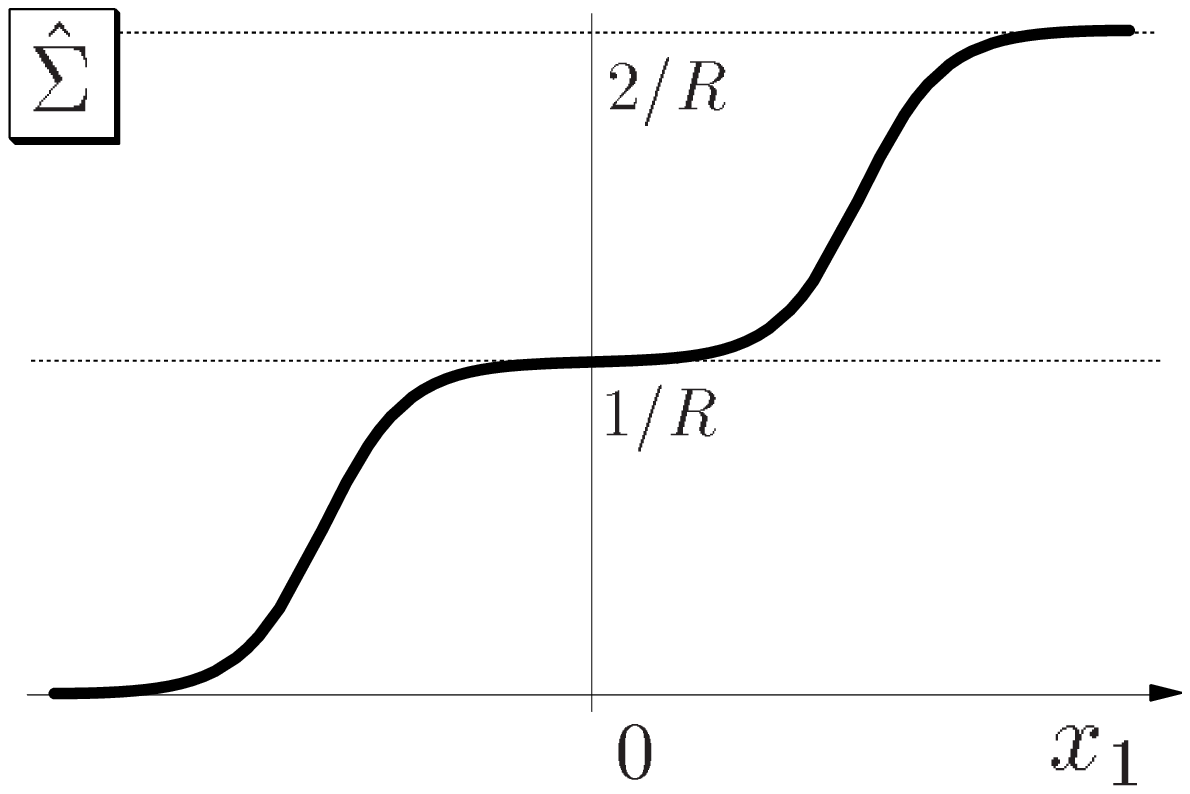} &
\includegraphics[width=50mm]{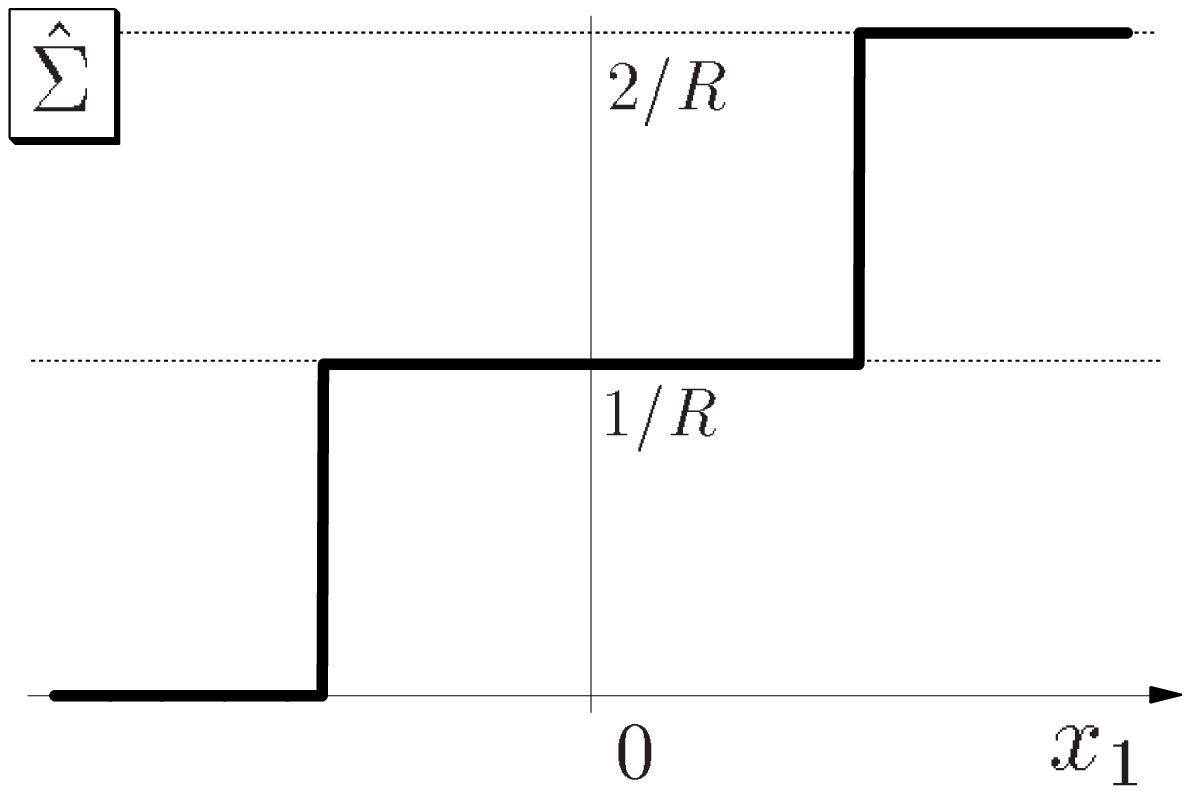} \\
(a) energy density & (b) $\tr \, \hat \Sigma$ & (c) $ \displaystyle \lim_{g\rightarrow \infty} \tr \, \hat \Sigma = \p_x N_P(x)$
\end{tabular}
\caption{(a) represents the energy density of two vortices. The energy is localized around the center of the vortices. (b) shows the profile of a kink solution, or equivalently $\tr \, \hat \Sigma$ as defined in \eqref{eq:1dim-Sigma}. In the strong gauge coupling limit, the profile reduces to step-wise shape as shown in (c).}
\label{fig:1dim-Sigma}
\end{center}
\end{figure}

If we dimensionally reduce the theory on $S^1$, 
then the vortex can be viewed as a domain wall 
in (1+1)-dimensional theory \cite{Isozumi:2004jc}. 
These field theoretical BPS solitons are realized 
by kinky D-brane configurations in superstring 
theory \cite{Lambert:1999ix,Eto:2004vy}, 
and the relation between vortices and domain walls 
is understood via T-duality \cite{Eto:2006mz}. 
In field theory language, 
the profile of the kink solution of the domain wall is described by 
a logarithm of a Wilson line along $S^1$ 
\beq
\hat \Sigma(x) \equiv - \frac{1}{2 \pi i R} 
\log \left[ \mathbf P \exp \left( i \int_{S^1} dy \, W_y \right) \right].
\label{eq:1dim-Sigma}
\eeq
$\hat \Sigma(x)$ can also be viewed as 
the adjoint scalar in the T-dual (dimensionally reduced) theory.
See Fig.~\ref{fig:1dim-Sigma}(b) for an example 
of a ${\rm Tr} \, \hat \Sigma(x)$ plot.

It is convenient to define a function $N_P(x)$ associated 
with the Laurent polynomial $P(u) = \det H_0(z)$ by
\beq
N_P(x) &\equiv& \lim_{g \rightarrow \infty} \int_0^{2\pi R} 
\frac{dy}{2\pi R} \, \frac{1}{2} \log \det \Omega \notag \\
&=& \int_0^{2\pi R}  \frac{dy}{2\pi R} \, \log |\det H_0| \notag \\
&=& \frac{1}{R} \sum_{i=1}^k \Big(x \, \theta(x-x_i) + x_i \theta(x_i - x) \Big).
\label{eq:1dim-Ronkin}
\eeq
In the final line, we have used the Jensen's 
formula \cite{Jensen}\footnote{The classical Jensen 
formula states that for arbitrary holomorphic function $f(x)$ with zeros at 
$a_i\ (i=1,2,\ldots k)$, we have
\beq
\frac{1}{2\pi}\int_0^{2\pi} \log |f(r e^{i\theta})| d\theta
=\log|f(0)|+\sum_{i=1}^{N_r} \log |\frac{r}{a_i}|,
\eeq
where we have chosen indices $i$ such that 
$a_i<r$ 
for $i=1,2,\ldots N_r$ and 
$a_i>r$ otherwise.
}.
As we will see in the next section, 
this piece-wise linear function is 
the ``Ronkin function'' in one dimension. 
By using this function $N_P$, 
the trace of the adjoint scalar 
in the infinite gauge coupling limit can be written by a step function
\beq
\lim_{g \rightarrow \infty} \tr \, \hat \Sigma ~=~ \p_x N_P(x) ~=~ \frac{1}{R} \sum_{i=1}^k \theta(x - x_i), \hs{10} x_i \equiv R \log |u_i| = {\rm Re} \, z_i.
\label{eq:g-inf-Sigma}
\eeq
Note that in the strong gauge coupling limit, 
the smooth kinky profile reduces to step-wise 
profile\footnote{
This limit is different from the one taken in \cite{Eto:2006mz} 
where the profile is not step function but 
has a constant slope in the interval of a vortex size.
} 
as shown in Fig.~\ref{fig:1dim-Sigma} (c). 

We can also express topological charges 
in terms of $N_P$ and its derivative $\hat{\Sigma}$.
The BPS bound \eqref{eq:BPS-bound-2} is now rewritten as 
\beq
E \ge -c \int d^2x \, \tr \, F_{xy} = \hat c \int dx \, \p_x \tr \, \hat \Sigma, \hs{10} \hat c \equiv 2 \pi R c,
\label{eq:1dim-energy}
\eeq
where we have used $\int dy \, \p_y \tr \, W_x = 0$ 
and $\tr \, \hat \Sigma = - \frac{1}{2\pi R} \int dy \, \tr \, W_y$. 
In the context of the domain wall, 
the quantity $ \hat c \int dx \, \p_x \tr \, \hat \Sigma$ 
provides the sum of the charges (masses) of the domain walls 
\cite{Isozumi:2004jc}. 
Interestingly, the energy of the BPS configuration 
can be determined only from $\tr \, \hat \Sigma$, 
namely the zero mode of the overall $U(1)$ gauge field $\tr \, W_y$. 
This is because the topological charge is 
determined only from the boundary condition 
and all the massive KK-modes vanish at spatial infinities 
$x \rightarrow \pm \infty$ in the BPS configurations. 

We finally discuss the K\"ahler metric 
on the moduli space of BPS vortices.
The moduli space of the vortices is a K\"ahler manifold 
parametrized by the moduli parameters $\phi_i, \bar \phi_i$. 
The K\"ahler metric of the moduli space is 
directly calculated from the solution of the master 
equation as follows \cite{Eto:2006uw} 
(see also \cite{Eto:2004rz,Eto:2006pg})
\beq
K_{i \bar j} &=& c 
\int d^2 x \, \mathcal K_{i \bar j}
(z, \bar z, \phi, \bar \phi, R, l), \label{eq:1dim-Kahler1} \\
\mathcal K_{i \bar j} &\equiv&  
\tr \left[ \p_i \p_{\bar j} \log \Omega + 4 l^2 
\Big( 
\p_{\bar z} (\Omega \p_i \Omega^{-1}) 
\p_{\bar j} (\Omega \p_z \Omega^{-1})
- 
\p_{\bar z} (\Omega \p_z \Omega^{-1}) 
\p_{\bar j} (\Omega \p_i \Omega^{-1})  \Big) \right], 
\label{eq:1dim-Kahler2}
\eeq
where 
$\p_i \equiv \p/\p \phi_i,~\p_{\bar j} \equiv \p/\p \bar \phi_j$ 
are derivatives with respect to the moduli parameters 
and $l \equiv 1/g \sqrt{c}$ is the length scale of the vortex core. 
Note that the matrix-valued 
function $\Omega$ depends on 
the parameters $g$ and $c$ only through $l$. 

For concreteness, let us consider $k$-vortex configurations 
in the Abelian-Higgs model ($\NC=\NF=1$). 
In this case the $2k$-dimensional moduli space is 
parameterized by the positions of vortices $z_i~(i=1,\cdots,k)$. 
For well-separated vortices $|z_i - z_j| \gg R,\, l$, 
the asymptotic metric is obtained by taking the limit 
$l \rightarrow 0,\, R \rightarrow 0$. 
In the small vortex limit $l \rightarrow 0$ the K\"ahler metric becomes
\beq
K_{i \bar j} \approx c \int d^2 x \frac{\p^2}{\p z_i \p \bar z_j} \log |H_0|^2.
\label{eq:1dim-asymptotic-metric}
\eeq
Therefore the K\"ahler potential $K$, which determines the metric by a relation 
$K_{i \bar j} = \frac{\p^2 K}{\p z_i \p \bar z_j}$, 
can be written by
\beq
K \approx 4 \pi c \int dx \, \Big( F_P(x,z_i,\bar z_i) - f(x,z_i) - \overline{f(x, z_i)} \Big),
\label{eq:1dim-asymptotic-KP}
\eeq
where $F_P(x) \equiv \lim_{R \rightarrow 0} R N_P(x)$ 
and $f(z_i)$ is a holomorphic function 
which is required to make the K\"ahler potential finite. 
This $F_P$ is a one-dimensional ``tropical polynomial'' 
that we will extend to the two-dimensional case in the next section. 
Since the asymptotic forms of the function $F_P(x)$ are given by
\beq
F_P(x) = \left\{ \begin{array}{cc} x_1 + \cdots + x_k,&~~\, x \rightarrow - \infty \\ k x, & x \rightarrow \infty \end{array} \right.,
\label{eq:1dim-Fp}
\eeq
a possible choice of the function $f(x,z_i)$ is
\beq
f(x,z_i) = \frac{z_1 + \cdots + z_k}{2} \theta(-x) + \frac{k x}{2} \theta(x).
\label{eq:couter-term}
\eeq
\begin{figure}[h]
\begin{center}
\includegraphics[width=60mm]{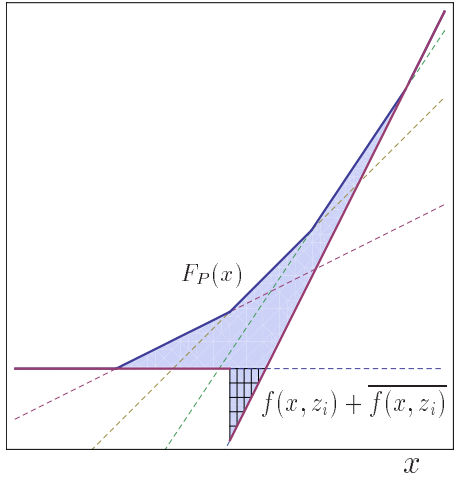}
\caption{The asymptotic K\"ahler potential can be evaluated 
as the area of the region surrounded by $F_P(x)$ and 
$f(x,z_i) + \overline{f(x,z_i)}$ (shaded regions). 
The area of the meshed region gives the contribution from the center of mass modulus $\frac{\pi c k}{2} (z_c + \bar z_c)^2,~ z_c \equiv (z_1 + z_2 + \cdots + z_k)/k$.}
\label{fig:1dimKP}
\end{center}
\end{figure}
Then the asymptotic K\"ahler potential 
can be evaluated as the area of the shaded region in Fig.\,\ref{fig:1dimKP}, 
which is given by 
\beq
K \approx 2 \pi c \sum_{i=1}^k x_i^2 = \frac{\pi c}{2} \sum_{i=1}^k (z_i + \bar z_i)^2.
\label{eq:flat-KP}
\eeq
Using this K\"ahler potential, the K\"ahler metric is given 
by $K_{i \bar j} = \p_i \bar \p_j K = \pi c \, \delta_{i \bar j}$. 
So the effective Lagrangian which describes the dynamics of 
well-separated vortices becomes 
\beq
L_{\rm eff} = K_{i \bar j} \dot z_i \dot{\overline{z}}_j = \pi c \sum_{i=1}^k |\dot z_i|^2.
\eeq
This shows that the well-separated vortices 
behave as undistinguished free particles with the mass $2 \pi c$.

In the next section, 
we move on to the case of vortex on $(\C^\ast)^2$. 
Although the story is more complicated and general, 
we will encounter similar structures to those described in this section.

\section{Webs of Vortex Sheets on $(\C^\ast)^2$
}\label{sec:vortex_sheets_C2}
\subsection{Vortex Sheets and Amoeba}\label{sec:amoeba}
Let us consider the vortex-instanton system 
on $(\C^\ast)^2 \simeq \R^2 \times T^2$. 
As before, we will use $(x_1,\,y_1,\,x_2,\,y_2)$ 
and $z_1 \equiv x_1 + i y_1 ,~z_2 \equiv x_2 + i y_2$ 
as real and complex coordinates of $(\C^\ast)^2$, respectively. 
The coordinates of $T^2$ are identified 
with periods $(2\pi R_1,\,2\pi R_2)$, namely 
$y_i \sim y_i + 2 \pi R_i$. In this case, 
the determinant of the moduli matrix $\det H_0$, 
which defines the vortex sheets, 
is written in the form of the Fourier series
\beq
\det H_0(z_1,z_2) = \sum_{(n_1,n_2) \in \mathbb Z^2} a_{n_1,n_2} \, e^{ \frac{n_1}{R_1} z_1 + \frac{n_2}{R_2} z_2}.
\label{eq:2dim-Fourier}
\eeq 
If we define new cylindrical coordinates $(u_1, u_2)$ 
on $(\mathbb C^\ast)^2$ by $u_i \equiv e^{\frac{z_i}{R_i}}$, 
$\det H_0$ is now written by a Laurent polynomial
\beq
P(u_1,u_2) ~~\equiv~~ \det H_0 ~~= \sum_{(n_1,n_2) \in \mathbb Z^2} a_{n_1,n_2} \, u_1^{n_1} u_2^{n_2}.
\label{eq:2dim-LP}
\eeq
The positions of the vortices are described by zeros of $P(u_1,u_2)$ 
similarly to those on $\C^\ast$ in the previous section, 
but the vortices form a two-dimensional sheet (surface) 
in $(\C^\ast)^2$ in the present case. 

We define the ``Newton polytope'' $\Delta(P) \subset \R^2$ of a Laurent polynomial $P(u_1,u_2)$ by
\beq
\Delta(P) = {\rm conv.\ hull} ~ \left\{\left(n_1, n_2 \right) \in \mathbb Z^2 \Big|~ a_{n_1,n_2} \not = 0 \right\}.
\label{eq:Newton}
\eeq
Conversely, $P(u_1,u_2)$ is called the Newton polynomial of $\Delta$, 
when its Newton polytope $\Delta$ is convex. 
In the discussion of domain wall webs,  
the Newton polytope 
$\Delta(P)$ was called the ``grid diagram" 
\cite{Eto:2005cp}--\cite{Eto:2007uc}. 
When we say ``Newton polynomial'', 
the coefficients $a_{n_1,n_2}$ in \eqref{eq:2dim-LP} 
are arbitrary parameters. 
Namely, $a_{n_1,n_2}$ are regarded as moduli parameters of the vortices. 

\begin{figure}[htbp]
\begin{center}
\begin{tabular}{ccc}
\includegraphics[width=60mm]{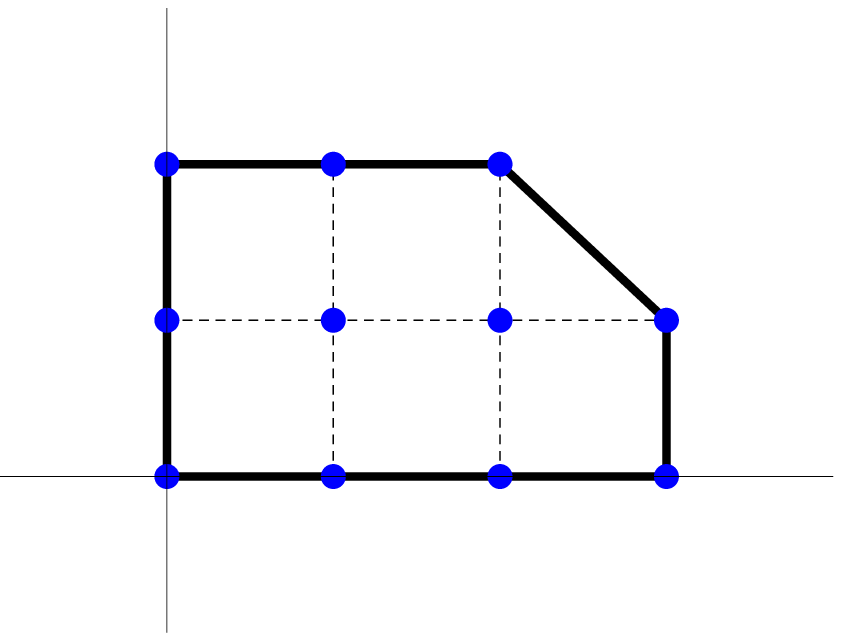} & \hs{5} &
\includegraphics[width=50mm]{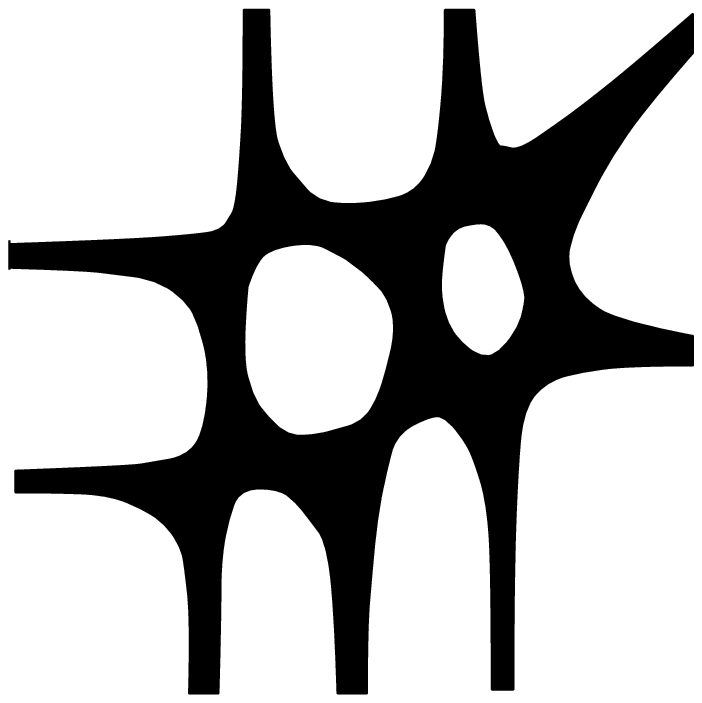} \\
(a) Newton polytope & & (b) amoeba
\end{tabular}
\end{center}
\caption{An example of amoeba; $P(u_1,u_2) = a_{0,0} 
+ a_{1,0} u_1 + a_{2,0} u_1^2 + a_{3,0} u_1^3 + a_{0,1} u_2 
+ a_{1,1} u_1 u_2 + a_{2,1} u_1^2 u_2 + a_{3,1} u_1^3 u_2 
+ a_{0,2} u_2^2 + a_{1,2} u_1 u_2^2 + a_{2,2}u_1^2 u_2^2$.}
\label{fig:amoeba}
\end{figure}

Analogous to  the case of the vortices on the cylinder 
discussed in the the previous section, 
a web of vortices on $(\mathbb C^\ast)^2$ 
is now dimensionally reduced 
to a web of domain walls on $\mathbb \R^2$ 
\cite{Eto:2005cp}--\cite{Eto:2007uc}. 
In order to see the connection better, 
we define ``amoeba'' of $P$ by\footnote{Amoeba can be defined for $(\C^\ast)^n$ with arbitrary integer $n$, but we only use the case of $n=2$.}
\beq
\mathcal A_{P} = \Big\{ \big( R_1 \log |u_1|,~R_2 \log|u_2|\big) \in \mathbb R^2 ~\big|~ P(u_1,u_2) = 0 \Big\}.
\label{eq:amoeba}
\eeq
Note here that $ R_1 \log|u_1| = x_1$ and $R_2 \log|u_2| = x_2$.
This is a projection of the shape of vortex sheet onto 
two non-compact directions. 
See Fig.~\ref{fig:amoeba} for an example of amoeba. 
From this example, we can see that the amoeba has several 
asymptotic regions extending to infinity, 
which are called ``tenticles'' in literature. 
In physics terms, each of these tenticles represents 
a semi-infinite cylinder of the vortex, 
and, for generic coefficients, corresponds 
to the normals to the Newton polytope (see Fig.~\ref{fig:amoeba}).\footnote{When moduli parameters are chosen to be special value, several tenticles of amoeba can merge into one. In this case, the multiplicity of the spires is considered to be greater than one.}
We also have (again, for generic values of moduli) holes 
for each internal lattice point of the Newton polytope. 
We learn from these facts that amoeba is 
a projection of generic webs of the vortices.

The notion of amoeba dates back to \cite{GKZ}. 
It was originally studied in the context of 
monodromy of the so-called GKZ-hypergeometric 
(originally called $\mathcal{A}$-hypegeometric) functions. 
It is also intimately connected with real algebraic geometry 
(Hilbert's 16th problem). 
Furthermore, it plays an important role 
in the discussion of the tropical geometry, as we will see. 
What we have found is that amoeba also appears quite naturally 
in the discussion of the webs of the vortices. 
See also the work \cite{Maeda:2006we}, 
which discusses amoeba in the context of instanton counting.

To see the relation between the webs of the vortex sheets 
and the webs of the domain walls, it is convenient (just as in 
the previous section) 
to define $\hat \Sigma_1(x_1,x_2)$ by
\beq
\hat \Sigma_1(x_1,x_2) \equiv - \frac{1}{2\pi R_1} 
\oint \frac{dy_2}{2\pi i R_2} \log \left[ 
\mathbf P \exp \left( i \oint dy_1 \, W_{y_1} \right) \right],
\label{eq:2dim-Sigma}
\eeq
and similarly for $\hat \Sigma_2(x_1,x_2)$ 
by interchanging the subscript 1 with 2. 
These adjoint scalar fields $\hat \Sigma_i(x_1,x_2)$ on $\R^2$ 
are interpreted as the zero modes of the gauge fields 
in the Kaluza-Klein decomposition and 
exhibit the two-dimensional kink profiles. 
The trace of these adjoint scalar fields can be simply written as
\beq
\tr \left[ \hat \Sigma_i(x_1,x_2) \right] 
&=& - \frac{1}{2\pi R_1} \frac{1}{2\pi R_2} \int d^2y \, 
\tr \, W_{y_i} \notag \\
&=& \frac{1}{8\pi^2 R_1 R_2} \frac{\p}{\p x_i} \int_{T^2} d^2y \log \det \Omega,
\label{eq:Sigma}
\eeq 
where we have used Eq.\,(\ref{eq:gauge-S}) and $\Omega=SS^\dagger$. 
The matrix-valued function $\Omega$ is simplified 
in the strong gauge coupling limit to 
(recall master equation \eqref{eq:master}) 
\beq
\lim_{g \rightarrow \infty} \Omega~=~\Omega_0 ~=~ \frac{1}{c} H_0 H_0^\dagger.
\label{eq:Omegainfinity}
\eeq
Although the vortex sheet becomes thin and singular 
in the strong coupling limit, 
the matrix $\Omega_0$ still has important physical informations 
on vortices and instantons. 
The traces of the adjoint scalar fields Eq.\,(\ref{eq:Sigma}) 
in the strong coupling limit are given by
\beq
\lim_{g \rightarrow \infty} \tr \left[ \hat \Sigma_i(x_1,x_2) \right] &=& \frac{\p}{\p x_i} N_{P}(x_1,x_2), 
\label{eq:2dim-g-inf-Sigma}
\eeq
where $N_{P}(x_1,x_2)$ is nothing but the ``Ronkin function" 
\cite{Ronkin} in two dimensions
\beq
N_{P}(x_1,x_2) &=& \frac{1}{2\pi R_1} \frac{1}{2\pi R_2} 
\int_{T^2} d^2y \, \log |\det H_0(z_1,z_2)| \nonumber \\
&=& \frac{1}{(2 \pi i)^2}  
\int_{|u_i|=
e^{x_i/R_i}
} 
\frac{du_1}{u_1} \wedge \frac{du_2}{u_2} \, \log |P(u_1,u_2)|,
\label{eq:Ronkin}
\eeq
defined from the Laurent polynomial $P(u_1,u_2)=\det H_0(z_1,z_2)$.

The Ronkin function has several interesting properties. 
First of all, it is convex \cite{Ronkin}. 
Second, the derivatives of the Ronkin function, 
$\Tr\ \hat{\Sigma}_1$ and $\Tr\ \hat{\Sigma}_2$, 
take constant values in each 
complement of the amoeba,\footnote{In \cite{ForsbergPassareTsikh} and many other literature, these constant values are called the orders of the complement of the amoeba.}
and those constant values (multiplied by $R_1 R_2$) 
are given by the lattice points 
in the Newton polytope of $P$ \cite{ForsbergPassareTsikh}. 
More generally, $R_1 R_2 ( \tr \, \hat{\Sigma}_1, \tr \, \hat{\Sigma}_2)$ 
as a function defined on $\mathbb{R}^2$ (including points on the amoeba)
take values within the Newton polytope $\Delta(P)$ of $P$.

\subsection{Relation with Tropical Geometry}
\label{sec:tropical_geomegtry}

Now one difference arises from the previous section. 
In the complex one-dimensional case 
discussed in the previous section, 
the Ronkin function $N_P$ is piece-wise linear 
when we take the thin wall limit $l=1/ g \sqrt{c} \to 0$. 
In two-dimensional case, however, 
the Ronkin function and its derivative are smooth 
even when the gauge coupling goes to infinity. 

We can still consider another limit 
in which the derivative of 
the Ronkin function becomes discontinuous. 
The limit is $R_1 = R_2 = R \rightarrow 0$ 
with fixed 
\begin{equation}
r_{n_1,n_2} \equiv R \log |a_{n_1,n_2}|
. 
\label{eq:ronkin_const}
\end{equation}
This limit corresponds to dimensionally reducing the theory 
to (2+1)-dimensions, neglecting all KK modes. 
In this limit, the amoeba degenerates into a set of lines (``spines''), 
which is called ``tropical variety'' in the tropical geometry literature. 
Physically speaking, the vortices reduce 
to the domain walls by dimensional reduction, 
and the tropical variety signifies the location of the domain walls. 
At the same time, the Ronkin function becomes
a piece-wise linear function $F_{P}(x_1,x_2)$ defined by
\beq
F_{P}(x_1,x_2) ~=~ \lim_{R \rightarrow 0} R \, \log |P(u_1,u_2)| ~=~ \underset{(n_1,n_2) \in V(Q)}{\rm max} \left( n_1 x_1 + n_2 x_2 + r_{n_1,n_2} \right),
\label{eq:2dim-FP}
\eeq
where $V(Q)$ is a set of the vertices associated with the Newton polytope $Q$, 
and $r_{n_1,n_2}$ in Eq.(\ref{eq:ronkin_const}) are constants 
determined from constants $a_{n_1,n_2}$ in \eqref{eq:2dim-LP}. 
This function $F_P$ coincides with the Ronkin function $N_P$ 
on the complement of the amoeba 
(recall $N_P$ is linear on each complement).

If we compare \eqref{eq:2dim-FP} with \eqref{eq:2dim-LP}, 
we notice that the sum and products in the polynomial 
$\sum a_{n_1,n_2} u_1^{n_1} u_2^{n_2}$ are replaced 
by a maximum function  
$\underset{(n_1,n_2) \in V(Q)}{\rm max} \left( n_1 x_1 + n_2 x_2+ r_{n_1,n_2} \right)$ 
of the linearized functions. 
The formal reasoning is given as follows. 
If we define $\tilde{x}_1=\exp(x_1/R)$, $\tilde{x}_2=\exp(x_2/R)$,
$\tilde{x}_1+\tilde{x}_2=\exp(x_3/R)$, and 
$\tilde{x}_1 \tilde{x}_2=\exp(x_4/R)$, then we find
\beq
x_3={\rm max}(x_1,x_2),\ \  x_4=x_1+x_2
\eeq
in the $R\to \infty$ limit. 
Hence in the tropical limit, 
the ring $(\mathbb{R},+,\times)$ is replaced by 
an idempotent semiring\footnote{A semiring is an algebraic 
structure similar to a ring, but without the requirement 
that each element must have an additive inverse.} 
$(\mathbb{R},\oplus,\otimes)$, 
with a tropical addition $\oplus$ and 
a tropical multiplication $\otimes$ given by
\beq
x_1 \oplus x_2={\rm max}(x_1,x_2),\ \ x_1\otimes x_2=x_1+x_2,
\eeq
respectively.
The semiring $(\mathbb{R},\oplus,\otimes)$ 
is sometimes called the tropical semiring or the max-plus algebra. 

The operation replacing the addition and multiplication 
with the tropical addition and tropical multiplication 
is also called dequantization or ultradiscretization. 
It appears in a discretization of integrable soliton equations 
such as KdV, Toda and KP hierarchies and also in cellular automata. 
These integrable soliton systems seem to be completely different 
from the vortex-instanton system we are considering, 
but it is interesting that the same structure 
plays important roles in many integrable systems.

We have mentioned about the tropical limit and tropical semiring, but then what is the corresponding geometry?
In usual algebraic geometry, 
we consider geometry corresponding to commutative ring. 
In contrast, the geometry corresponding to 
tropical semiring $(\mathbb{R},\oplus,\otimes)$ 
is called {\em tropical (algebraic) geometry}.\footnote{According to \cite{SpeyerSturmfels}, the name ``tropical'' was coined by a French mathematician Jean-Eric Pin \cite{Pin}, in honor of their Brazilian colleague Imere Simon \cite{Simon}.}
We can formulate and prove ``tropical analogue'' 
of many theorems in usual algebraic geometry, 
such as the Riemann-Roch theorem and the Bezout's theorem. 
Although the study of idempotent semirings 
in applied mathematics (such as control theory and optimization) 
has a long history \cite{Control}, 
the study of corresponding geometry is 
relatively new and it is still an active area of research 
(see \cite{FirstSteps,SpeyerSturmfels,MikhalkinICM}). 
The tropical geometry has now diverse applications, 
ranging from 
string networks \cite{Ray:2008xq}, 
enumeration of curves \cite{MikhalkinEnumerative}, 
mirror symmetry \cite{Abouzaid} and even computational biology \cite{biology}.

\begin{figure}[h]
\begin{center}
\begin{tabular}{ccc}
\includegraphics[width=50mm]{amoeba2.eps} & ~\raisebox{1.5cm}{$\overrightarrow{\hs{8} R \rightarrow 0 \phantom{\Big[} \hs{7}}$}~ &
\includegraphics[width=50mm]{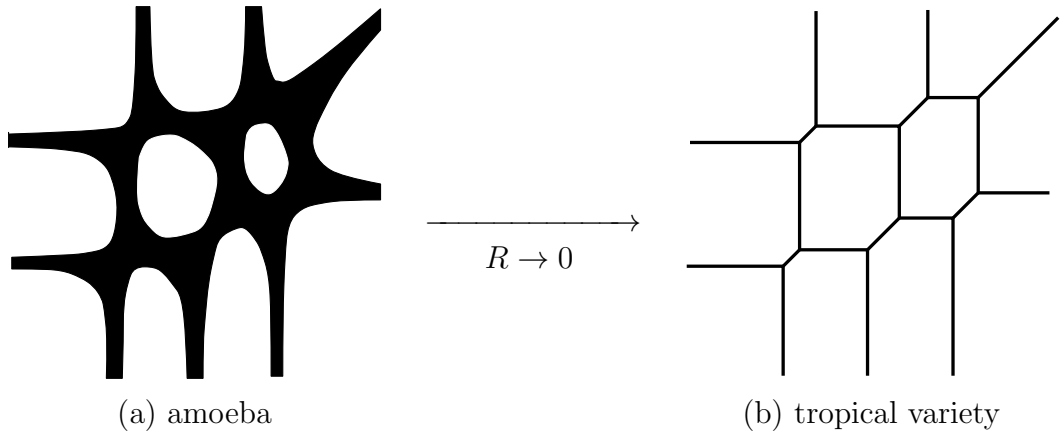} \\
(a) amoeba & & (b) tropical variety
\end{tabular}
\end{center}
\caption{An example of the amoeba and corresponding tropical variety.}
\label{fig:tropvar}
\end{figure}

We can consider a tropical version of algebraic variety, 
namely tropical variety. 
In the literature, it is often defined as a non-Archimedian amoeba, 
but for our applications, 
it suffices to define it as the set of points 
where the piece-wise linear function 
$F_P(x_1,x_2)$ (``tropical polynomial'') is not differentiable. 
This is nothing but the skeleton (spine) of the amoeba in the limit $R\to 0$, 
and its physical meaning is the position of the domain walls, 
namely the position of the step-wise kinks appearing 
in the profiles of $\tr \, \hat \Sigma_i(x_1,x_2)~(i=1,2)$.
An example of the tropical varieties are shown in Fig.~\ref{fig:tropvar}. 
As shown there, tropical varieties (in the situation we want to consider) 
are obtained from triangulation of the Newton polytope 
(\cite{FirstSteps}, Proposition 3.5). 
In this sense, this is similar to the so-called 
$(p,q)$-web or web diagram in \cite{Aharony:1997bh}. 
We will make more comments on this analogy 
in the last section devoted to the discussion. 

\begin{figure}[htbp]
\begin{center}
\begin{tabular}{ccc}
\includegraphics[width=45mm]{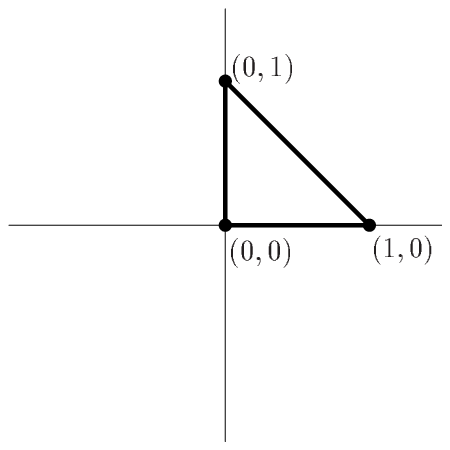} &
\includegraphics[width=45mm]{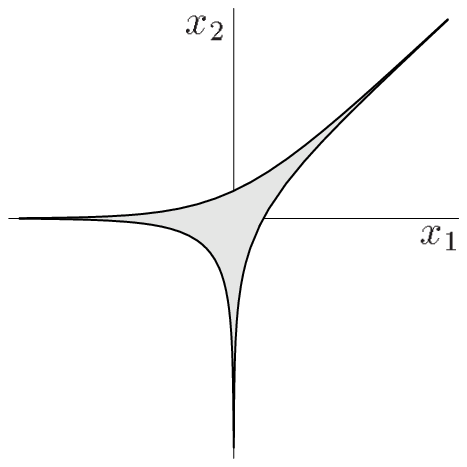} &
\includegraphics[width=45mm]{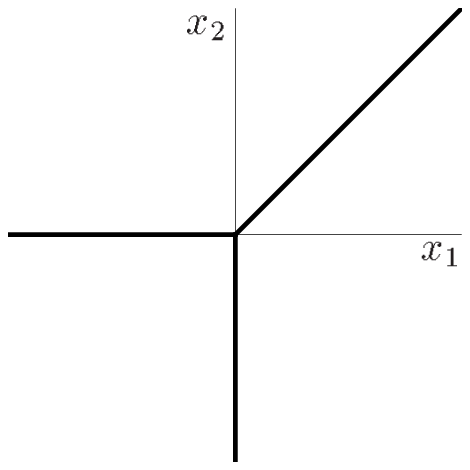} \\
(a) Newton polytope & (b) amoeba & (c) tropical variety
\end{tabular}
\end{center}
\caption{The Newton polytope (a) amoeba (b) and tropical variety (c) for 
the Laurent polynomial $P(u_1,u_2) = u_1 + u_2 + 1$.}
\label{fig:Ronkin-eg}
\end{figure}

\paragraph{Example\\}
Let us here give a simple and concrete example for later discussions. 
An example of the Newton polytope is given in Fig.~\ref{fig:Ronkin-eg} (a) and we set $R_1=R_2=1$ for simplicity in the following. 
Then the corresponding Newton polynomial is given by
\beq
P(u_1,u_2) = u_1 + u_2 + 1 = e^{z_1} + e^{z_2} + 1,
\label{eq:example-P}
\eeq
and its amoeba and corresponding variety is shown in Fig.~\ref{fig:Ronkin-eg}.

\begin{figure}[htbp]
\begin{center}
\begin{tabular}{ccc}
\includegraphics[width=60mm]{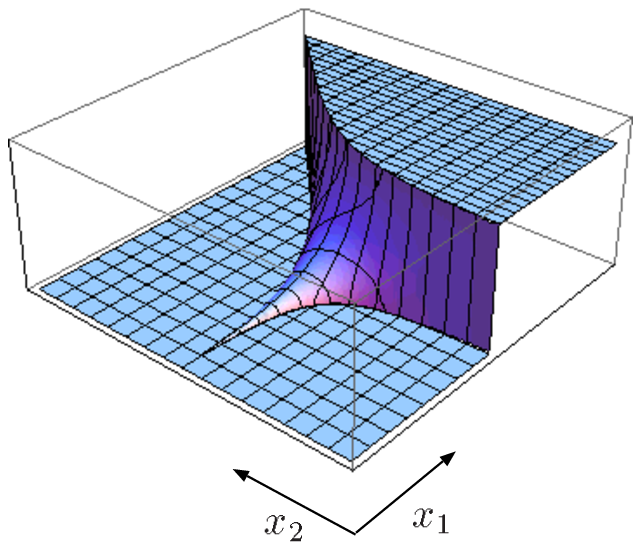}& \hs{10} &
\includegraphics[width=60mm]{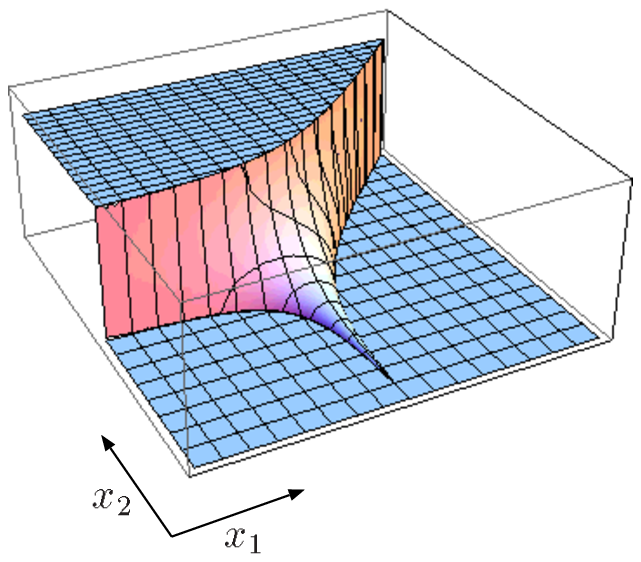}\\
$\displaystyle \lim_{g\rightarrow \infty} \tr \, \hat{\Sigma}_1$ \hs{5} & &
$\hs{5} \displaystyle \lim_{g\rightarrow \infty} \tr \, \hat{\Sigma}_2$
\end{tabular}
\end{center}
\caption{The plots of $\tr \, \hat{\Sigma}_1$ and $\tr \, \hat{\Sigma}_2$.}
\label{fig:Ronkin-plot}
\end{figure}

The derivatives Tr \, $\hat{\Sigma}_i(x_1,x_2)$ of the 
Ronkin function associated with $P$ are computed to be 
\beq
\lim_{g \rightarrow \infty} \tr \left[ \hat \Sigma_1(x_1,x_2) \right] = 
\left\{
\begin{array}{ccl}
0 & \text{for} & x_1 < \displaystyle \log \left| e^{x_2} - 1 \right| \\
\vs{2} \displaystyle 1 - \frac{1}{\pi}\cos^{-1} \left( \frac{e^{2 x_1} - e^{2 x_2} - 1}{2 e^{x_2}} \right) & \text{for} & \displaystyle \log \left| e^{x_2} - 1 \right| \leq x_1 \leq \displaystyle \log \left| e^{x_2} + 1 \right| \\
\displaystyle 1 & \text{for} & x_1 > \displaystyle \log \left| e^{x_2} + 1 \right|
\end{array}
\right.,
\nonumber
\label{eq:example-Sigma1}
\eeq
\beq
\lim_{g \rightarrow \infty} \tr \left[ \hat \Sigma_2(x_1,x_2) \right] = 
\left\{
\begin{array}{ccl}
0 & \text{for} & x_2 < \displaystyle \log \left| e^{x_1} - 1 \right| \\
\vs{2} \displaystyle 1 - \frac{1}{\pi}\cos^{-1} \left( \frac{e^{2 x_2} - e^{2 x_1} - 1}{2 e^{x_1}} \right) &  \text{for} & \displaystyle \log \left| e^{x_1} - 1 \right| \leq x_2 \leq \displaystyle \log \left| e^{x_1} + 1 \right| \\
\displaystyle 1 &  \text{for} & x_2 > \displaystyle \log \left| e^{x_1} + 1 \right|
\end{array}
\right. ,
\nonumber
\label{eq:example-Sigma2}
\eeq
and their plots are given in Fig. \ref{fig:Ronkin-plot}. 
Note that $\hat{\Sigma}_i$ takes 
a constant value at each 
complement of amoeba, as expected.

\begin{figure}[htbp]
\begin{center}
\begin{tabular}{ccc}
\includegraphics[width=60mm]{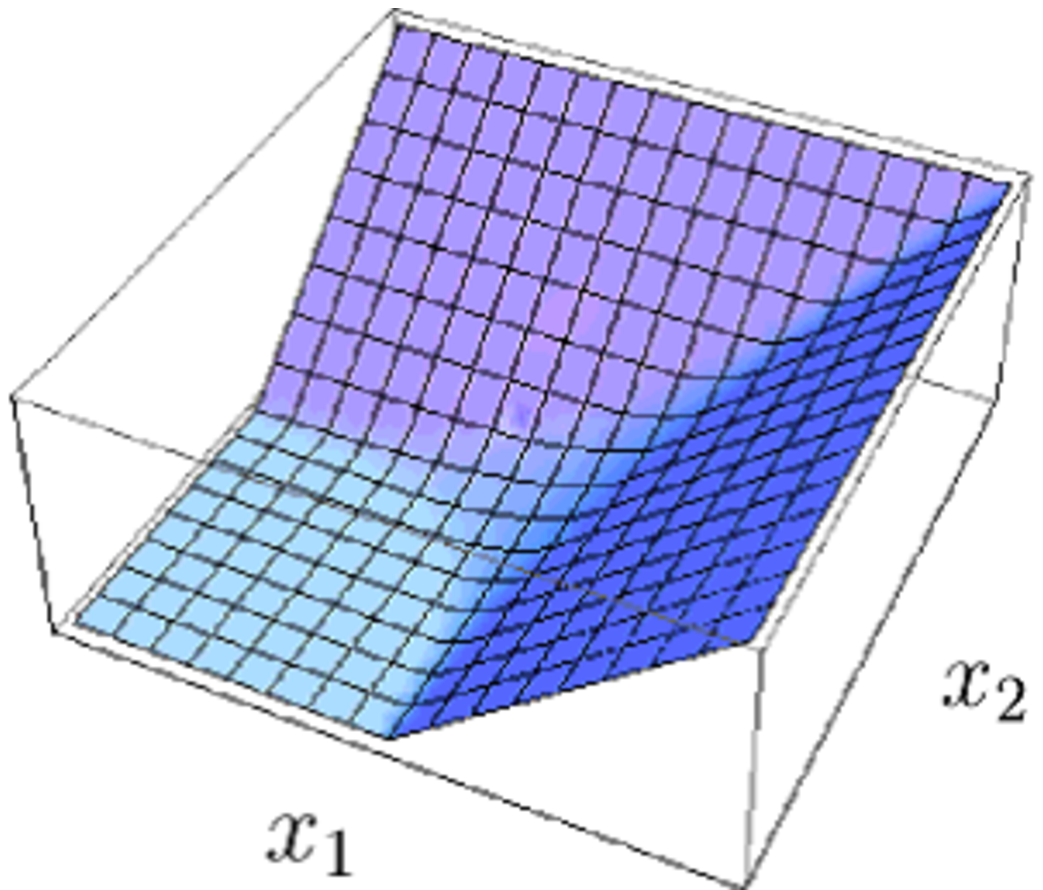} & \hs{10} &
\includegraphics[width=60mm]{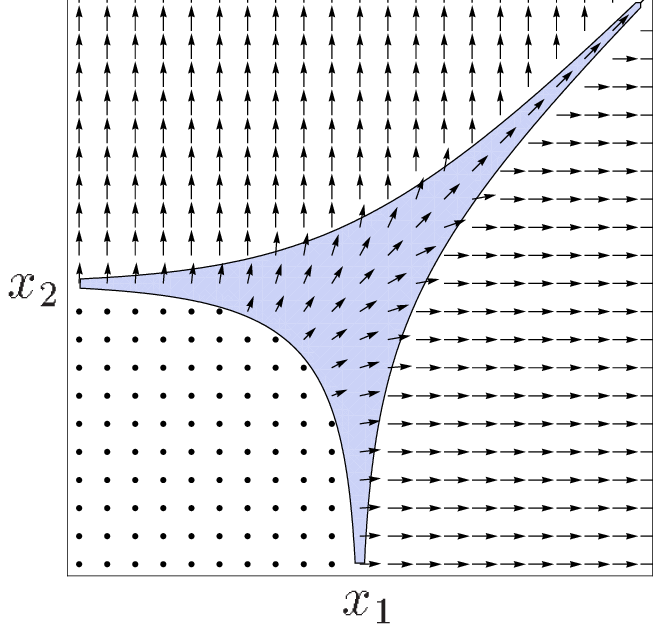} \\
(a) Ronkin function & & (b) gradient of Ronkin function 
\end{tabular}
\end{center}
\caption{(a) Ronkin function and (b) $\tr \, \hat{\Sigma}_1$ and $\tr \, \hat{\Sigma}_2$ as the gradient of the Ronkin function.}
\label{fig:Ronkin-plot2}
\end{figure}

\subsection{Topological Charges}\label{sec:topological_charge}

We now move on to discussion of the topological charges 
given in \eqref{eq:total-instanton} and \eqref{eq:vortex-charge}.
For a given Laurent polynomial $P$, these topological 
charges are evaluated as follows. 
First let us consider the vortex charge. 
Since the topological charges are independent of the gauge 
coupling constant $g$, we can take the strong gauge 
coupling limit $g \rightarrow \infty$. 
In the strong gauge coupling limit, the magnetic flux of 
the overall $U(1)$ can be written as 
\beq
- \frac{1}{2\pi} \tr \, F = \frac{1}{4\pi} d d_c 
\log \det \Omega ~\rightarrow~ \frac{1}{2\pi} d d_c \log |P|,
\label{eq:first-Chern}
\eeq
where $d_c \equiv -i(\p - \bar \p)$. 
By using the Poincar\'e-Lelong 
formula\footnote{
This formula is the generalization of the formula 
$d d^c \log |z|=2\pi\delta^{2}(z)dx \wedge dy$ 
with $\int dx dy \, \delta^{2}(z)=1$.
}
\beq
\int_{(\C^\ast)^2} \frac{1}{2\pi} d d_c 
\log |P| \wedge \alpha = \int_{X} \alpha, \hs{10} X 
= \left\{ (u_1,u_2) \in (\C^\ast)^2 ~\big |~ P(u_1,u_2) = 0 \right\},
\label{eq:Poincare-Lelong}
\eeq
we can show that the vortex charge can be evaluated as 
\beq
 V ~=~ - c \int_{(\C^\ast)^2} \tr \, F \wedge \omega &=& 2 \pi c \int_{X} \omega ~=~ 2 \pi c \, {\rm Area}(X).
\label{eq:vortex-charge-Area}
\eeq
From this computation, it is clear that the vortex charge is distributed on the surface of the vortex sheets $X$.
We also see that the vortex charge is uniformly distributed along all $X$, and the total vortex charge is given by the area of the vortex sheets multiplied by the tension $2 \pi c$. It is interesting to note that the same formula has appeared in mathematics literature (\cite{PassareRullgard}, Theorem 6 and \cite{Rullgard}). \\

We can also give another expression for the vortex charge, 
using the Ronkin function $N_P$.
In the tropical limit $R_1,R_2\to 0$, the vortex charge is given by 
\beq
 V ~=~ \hat c \int_{\R^2} d^2 x 
\left( \p_1 \tr \, \hat \Sigma_1 + \p_2 \tr \, 
\hat \Sigma_2 \right) ~=~ \hat c 
\int_{\R^2} d^2 x \left( \p_1^2 + \p_2^2 \right) N_P(x_1,x_2),
\label{eq:vortex-Ronkin}
\eeq
where $\hat c \equiv 4 \pi^2 R_1 R_2 c$. 
Namely, vortex charge is given by an integration of a 
Laplacian of the Ronkin function.
If we take the limit $R_1 = R_2 = R \rightarrow 0$, the amoeba becomes the tropical variety which can be interpreted as the web diagram of the domain walls. The tension of each wall can be calculated as follows. Since the integrand of Eq.\,(\ref{eq:vortex-Ronkin}) becomes $\left( \p_1^2 + \p_2^2 \right) F_P(x_1,x_2)$ in the small radius limit $R \rightarrow 0$, the tension of the domain wall is computed by integrating the Laplacian of the piece-wise linear function $F_P(x_1,x_2)$ along the line perpendicular to the wall. If the wall is located along the line $n_1 x_1 + n_2 x_2 + r = (n_1+p) x_1 + (n_2 + q) x_2 + r'$, then the tension is given by
\beq
T_{(p,q)} = \frac{\hat c}{R} \sqrt{p^2 + q^2}.
\label{eq:wall-tension}
\eeq

Next let us consider the intersection charge 
\eqref{eq:intersection}. 
By taking the strong gauge coupling limit, the intersection 
charge density ${\cal I}_{\rm intersection}$ becomes a 
complex Monge-Amp\`ere measure $(dd_c \log|P|)^2$ on 
$(\C^*)^2$ associated with a 
plurisubharmonic\footnote{
Monge-Amp\`ere measure is defined for arbitrary 
plurisubharmonic function. 
Here it suffices to know that $\log|P|$ is 
plurisubharmonic for arbitrary holomorphic function $P$.} 
function $\log|P|$, which is a higher 
dimensional generalization of the Laplace operator 
(see \cite{Pluri,Demailly1} for discussion on the complex 
Monge-Amp\`ere measure):
\beq
{\cal I}_{\rm intersection} = \frac{1}{8\pi^2} 
\tr \, F \wedge \tr \, F ~&\rightarrow& 
- \frac{1}{\pi^2} \det\left( 
\frac{\del^2 \log|P|}{\del u_i \del \bar{u}_j} \right) 
du_1 \wedge d\bar{u}_1 \wedge du_2 \wedge d\bar{u}_2 \nn \\
&=& ~\, \, \frac{1}{8\pi^2} d d_c \log |P| \wedge d d_c \log |P|.
\label{eq:intersection1}
\eeq
Then the intersection charge is evaluated 
again by using Poincar\'e-Lelong formula, 
\beq
I_{\rm intersection} =
\frac{1}{8\pi^2} \int d d_c \log |P| \wedge d d_c \log |P| ~=~ \frac{1}{4\pi} \int_{X} d d_c \log |P|,
\label{eq:intersection-diverge}
\eeq
but this naive evaluation is unfortunately divergent.
The divergence comes from the fact that the strong gauge coupling limit in the master equation \eqref{eq:master} is ill-defined when $\Omega=0$, since there appears $\Omega^{-1}$ in the master equation. 
In principle, if we can solve master equation for finite gauge coupling, we could safely obtain a correct value of the intersection charge, but that would be difficult in practice. Instead, we propose to regularize the divergence as follows.

Let $P_1$ and $P_2$ be distinguished Laurent polynomials associated with the same Newton polytope $\Delta(P)$ of $P$, and replace two $P$'s in \eqref{eq:intersection-diverge} by $P_1$ and $P_2$, respectively.
For generic Laurent polynomials $P_1$ and $P_2$, the intersection points of the zero sets of $P_1$ and $P_2$ are discrete points. 
Then we obtain
\beq
\frac{1}{8\pi^2} \int d d_c \log |P_1| \wedge d d_c \log |P_2| ~=~ \frac{1}{4\pi} \int_{X_1} d d_c \log |P_2| ~=~ \frac{1}{2} \#(X_1 \cdot X_2),
\label{eq:intersection2}
\eeq
where the surfaces $X_i\ (i=1,2)$ are defined 
by $P_i(z_1,z_2)=0$ 
and the number of intersection points are denoted as 
$\#(X_1 \cdot X_2)$. 

Thanks to Bernstein's theorem \cite{Bernstein},\footnote{This theorem is a generalization of the well-known Bezout's theorem. See \cite{Sturmfels2} for leisurely introduction to Bernstein's theorem.}
$\#(X_1 \cdot X_2)$ is independent of the choice of the Laurent polynomials $P_1,P_2$ 
 as long as $P_1$ and $P_2$ are generic, and is given by $2 {\rm Area}(\Delta)$. We thus find that the intersection charge $I_{\rm intersection}$ is evaluated to be equal to the area of the Newton polygon:
\beq
I_{\rm intersection}={\rm Area}(\Delta).
\eeq

The meaning of this regularization is now clear. 
The original expression \eqref{eq:intersection-diverge} is 
divergent essentially because it is a self-intersection number. 
We propose to regularize this by infinitesimally changing $P$, 
but with fixed boundary conditions at infinity.\footnote{The 
condition that $P_1$ and $P_2$ are Newton polynomial of the 
convex polytope $\Delta(P)$ is important. 
Otherwise the answer depends on the choice of $P_1$ and $P_2$. 
For example, if we take $P_{1,j}=z_1$, $P_{2,j}=z_1+1/j$, 
then $dd^c \log |P_{1,j}| \wedge dd^c\log |P_{2,j}|=0$ for 
all $j$. 
If we take instead $P_{3,j}=z_1+z_2/j$, 
then $dd^c
\log |P_{1,j}| \wedge dd^c \log |P_{3,j}|=\delta_0
\equiv \delta^2(z_1)\delta^2(z_2)dx_1\wedge dy_1\wedge
dx_2\wedge dy_2
$. 
And for $P_{4,j}=z_1+z_2^j$ (in a neighborhood of $0$), 
$dd^c \log |P_{1,j}| \wedge dd^c \log |P_{4,j}|=j\delta_0$. 
All these functions converge to the same $P=z_1$ in the 
$j\to \infty$ limit, but gives a different answer. 
We thank Alexander Rashkovski for providing us with this example.} 



Instead of invoking Bernstein's theorem, we can take more 
down-to-earth approach and the calculation goes as follows. 
This derivation is not independent from the previous 
argument and moreover not rigorous, but it has an advantage 
of clarifying the relation with the Ronkin function and 
real Monge-Amp\`ere measure.

First, it is reasonable to expect\footnote{Essentialy, we are again using Bernstein's theorem here for the rigorous argument.} that the intersection number does not change under replacements $P_1(u_1,u_2) \rightarrow P_1(|u_1| e^{i \theta_1} , |u_2| e^{i \theta_2})$ and $P_2(u_1,u_2) \rightarrow P_2(|u_1| e^{i \phi_1} , |u_2| e^{i \phi_2})$, as far as $\theta_1, \theta_2, \phi_1$ and $\phi_2$ are sufficiently generic:
\beq
\#(X_1 \cdot X_2) &=& \#(X_1(\theta_1,\theta_2) \cdot X_2(\phi_1,\phi_2)),
\label{eq:V1V2intersection}
\eeq
where
\beq
X_1(\theta_1,\theta_2) &=& \{ (z_1,z_2) \in (\C^\ast)^2 ~\big|~ P_1(|u_1| e^{i \theta_1}, |u_2| e^{i \theta_2})=0 \}, \\
 X_2(\phi_1,\phi_2) &=&\{ (z_1,z_2) \in (\C^\ast)^2 ~\big|~ P_2(|u_1| e^{i \phi_1}, |u_2| e^{i \phi_2})=0 \}.
\label{eq:intersection3}
\eeq
Certainly the intersection number \eqref{eq:V1V2intersection} 
might change if $\theta_1$ and $\theta_2$ are non-generic, 
but those special values do not contribute when we integrate 
over all $\theta_1$ and $\theta_2$. 
The same applies to $\phi_1$ and $\phi_2$. 
Therefore the intersection charge can be written as
\beq
& &\frac{1}{8\pi^2} \int_{(\C^\ast)^2} 
d d_c \log |P_1| \wedge d d_c \log |P_2|\nn\\
 &=& \frac{1}{8 \pi^2} \int \frac{d \theta_1}{2\pi} 
\frac{d\theta_2}{2\pi} 
\frac{d \phi_1}{2\pi} \frac{d\phi_2}{2\pi} \int_{(\C^\ast)^2} d d_c 
\log \left|P_1(|u_1| e^{i \theta_1}, |u_2| e^{i \theta_2})\right| 
\wedge d d_c \log 
\left|P_2(|u_1| e^{i \phi_1}, |u_2| e^{i \phi_2})\right| \nn \\
&=&\frac{1}{8 \pi^2} \int_{\R^2\times T^2} dd_c N_{P_1}(x_1,x_2) 
\wedge dd_c N_{P_2}(x_1,x_2) \nn \\
&=& \frac{1}{8 \pi^2} \left( \frac{i}{2} \right)^2 
\int_{\R^2\times T^2} dz_i \wedge d\bar{z}_j  
\frac{\p}{\p x_i}  \frac{\p}{\p x_j}  N_{P_1}(x_1,x_2) 
\wedge dz_k \wedge d\bar{z}_l  \frac{\p}{\p x_k}  
\frac{\p}{\p x_l}  N_{P_2}(x_1,x_2) \nn \\
&=&\frac{1}{8 \pi^2} \int_{\R^2\times T^2} dx_1 \wedge dy_1 
\wedge dx_2 \wedge dy_2 \ \ \epsilon_{ik} \epsilon_{jl} 
\frac{\p}{\p x_i}  \frac{\p}{\p x_j}  N_{P_1}(x_1,x_2) \ 
\frac{\p}{\p x_k}  \frac{\p}{\p x_l}  N_{P_2}(x_1,x_2) \nn \\
&=& R_1 R_2 \int_{\R^2} dx_1\wedge dx_2\  \mu_{\rm MA}(P_1,P_2), 
\label{eq:intersection4}
\eeq
where
\beq
\mu_{\rm MA}(P_1,P_2)=\frac{1}{2!}\ \epsilon_{ik}
\epsilon_{jl} \frac{\p}{\p x_i}  \frac{\p}{\p x_j}  
N_{P_1}(x_1,x_2) \ \frac{\p}{\p x_k}  \frac{\p}{\p x_l}  N_{P_2}(x_1,x_2)
\eeq
is known as a real Monge-Amp\`ere measure,\footnote{
More precisely, when $P_1\ne P_2$, this is called a real 
mixed Monge-Amp\`ere measure.}
which is a real analogue of the complex Monge-Amp\`ere 
measure we explained previously. 
Note that this is well-defined since $N_P$ is convex, as 
discussed previously. If we set $P_1,P_2\to P$, then 
\beq
I_{\rm intersection}&=& R_1 R_2 \int_{\R^2} dx_1
\wedge dx_2\  {\rm Hessian}\ (N_P) \\
&=& \int d (R_1 {\rm Tr} \hat{\Sigma}_1) 
\wedge d(R_2 {\rm Tr} \hat{\Sigma}_2) \\
&=& {\rm Area}(\Delta(P)).
\eeq
In the final line we used the fact 
that $R_1 R_2 ( \Tr \hat{\Sigma}_1, \Tr \hat{\Sigma}_2)$ 
takes values in the Newton polytope $\Delta(P)$, as 
explained previously.
Interestingly, this result, that the total integral 
(or ``mass'' in the standard mathematics literature) of 
the real Monge-Amp\`ere measure is given by 
${\rm Area}(\Delta(P))$, has appeared previously in the 
mathematics literature (\cite{PassareRullgard} Theorem 4).

From the above calculation, we have shown that the evaluation of the complex Monge-Amp\`ere measure reduces to the evaluation of the real Monge-Amp\`ere measure. Since the Ronkin function is described by the zero mode in the KK decomposition, we learn from this fact that contributions from KK modes, although present, are canceled out in the final expression. This is just the same as in the discussion in Sec.\,\ref{sec:higgs_phase}: topological charge is determined only from the boundary condition and all KK modes vanish at spatial infinity.

\begin{figure}[htbp]
\begin{center}
\includegraphics[scale=0.4]{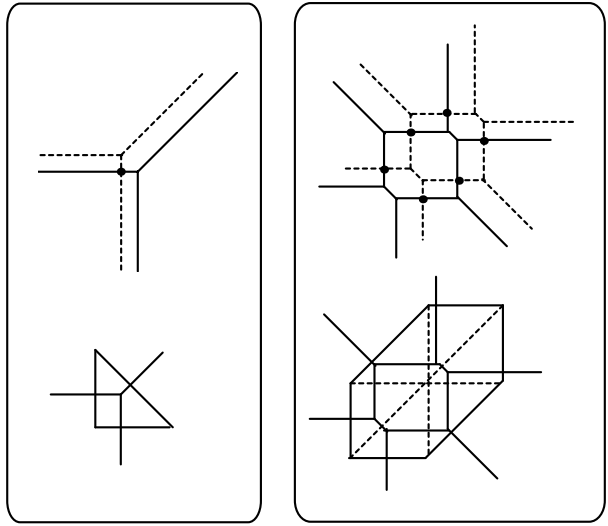}
\end{center}
\caption{Intersection of one tropical variety and its shift in generic directions. The corresponding Newton polytope $\Delta(P)$ is given below. It is easy to see that the number of intersection points is given by $2{\rm Area}(\Delta(P))$.}
\label{fig:tropicalintersection}
\end{figure}

We can provide one more different explanation of the formula 
$I_{\rm intersection}={\rm Area}(\Delta(P))$, which is much 
easier to understand (although strictly speaking, this is 
also just a restatement of the previous explanation). 
Taking the limit $R_1, R_2 \rightarrow \infty$,
the intersection 
charge $I_{\rm intersection}$ is given by the half of 
the intersection number of the tropical varieties of $P_1$ 
and $P_2$, and it is easy to see that the number is given 
by $2{\rm Area}(\Delta(P))$ 
(see Fig.~\ref{fig:tropicalintersection} for example). 
In tropical geometry, this statement is known as 
the tropical Bernstein theorem (\cite{Sturmfels} Theorem 9.5).

Although the computation here applies only to $R_1, R_2\rightarrow \infty$ limit, we expect that the intersection charge is still given by the same formula for finite $R_1, R_2$ as well, since the intersection charge is quantized and does not change continuously depending on $R_1, R_2$. For the same reason, although all the arguments so far are in the strong gauge coupling limit $g\to \infty$, we expect the same formula is kept in the finite gauge coupling as well.

So far, we have concentrated on the case of $\NC=\NF=N$ 
called local theory.
We expect that a 
similar formula holds even for $\NF > \NC$, 
which is often called semi-local theory.
For $\NC=1,\,\NF=2$, we can actually show this rigorously.\footnote{We thank Alexander Rashkovskii for providing us with the following proof.}
In the strong gauge coupling limit the solution of the master equation Eq.\,(\ref{eq:master}) for $H_0 =(P_1,P_2)$ is given by
\beq
\Omega \rightarrow \Omega_0 = |P|^2 \equiv |P_1|^2 + |P_2|^2.
\eeq
In this case, $I_{\rm intersection}$ needs no regularization. We use two representations of the current 
(intersection charge density) $[P]$ for $P=(P_1, P_2)$.
One is King's formula (a.k.a. the vector Poincar\'e, or the Poincar\'e-Martinelly formula \cite{Demailly1,GriffithsKing})
\beq
[P]=dd^c \log |P| \wedge dd^c\log |P|,
\eeq
and another is 
\beq
[P]=dd^c \log |P_1| \wedge dd^c \log |P_2|,
\eeq
which follows from Poincar\'e-Lelong formula applied to $P_2$ on $X_1$.
We thus have
\beq
I_{\rm intersection}&=&{1\over 8\pi^2} \int 
dd^c \log |P| \wedge dd^c\log |P|={1\over 8\pi^2} 
\int dd^c \log |P_1| \wedge dd^c\log |P_2|\\
&=&{\rm Area}(\Delta(P_1),\Delta(P_2)),
\eeq
where in the last line we have again used Bernstein's theorem, and ${\rm Area}(\Delta (P_1),\Delta (P_2))$ is a mixed volume defined by
\beq
2\,{\rm Area}(\Delta(P_1),\Delta(P_2))={\rm Area}(\Delta(P_1)+\Delta(P_2))-{\rm Area}(\Delta(P_1))-{\rm Area} (\Delta(P_2))
\eeq
and the sum (Minkowski sum) $\Delta(P_1)+\Delta(P_2)$ is defined by
\beq
\Delta(P_1)+\Delta(P_2) =\{x+y | x\in \Delta(P_1), y\in \Delta(P_2) \}
\eeq

For general $\NC$ and $\NF\,(>\NC)$, the solution of the master equation Eq.\,(\ref{eq:master}) in the strong coupling limit is given by
\beq
&\det \Omega ~\rightarrow~ \det \Omega_0 ~=~ |P|^2 \equiv \displaystyle \sum_i |P_i|^2,& \\
&P_i = \epsilon_{r_1 r_2 \cdots r_{\NC}} H_0^{r_1 A_1} H_0^{r_2 A_2} \cdots H_0^{r_{\NC} A_{\NC}}~~(i=1,\cdots,\NF!/\NC!(\NF-\NC)!).&
\eeq
Therefore the intersection charge $I_{\rm intersection}$ can be evaluated by integrating $(dd_c \log |P|)^2$ with $P$ being an $\NF!/\NC!(\NF-\NC)!$-dimensional vector.
Unfortunately, we have no mathematical estimate for 
$\NF\ge 3$, and only the upper bound \cite{Rashkovskii} 
for the integral of the complex Monge-Amp\`ere measure 
in $\C^2$ (not $(\C^\ast)^2$) is known. 
We conjecture that the result holds 
in this more general 
case in the same manner.\footnote{
At least, we can prove that intersection charge 
$I_{\rm intersection}$ is quantized. 
We thank Alexander Rashkovskii for providing us with 
this argument. The argument goes as follows. 
Let $P$ be a $N$-dimensional vector 
$\vec{P}=(P_1,P_2,\ldots, P_N)$. 
For a two-dimensional subspace $A$ of the $N$-dimensional 
complex space $(\C^\ast)^N$, let $P^A=(P_1^A,P_2^A)$ be 
$P_j^A=\sum_{i=1}^N\lambda_{ij}P_i$, where $\lambda_{ij}$ 
is the $2\times N$-matrix of orthonormal basis of $P$. 
Then by \cite{Demailly1,Demailly2}, Monge-Amp\`ere measure 
is represented as 
$$
(dd_c \log |P|)^2=\int_{G(N,2)} (dd_c \log |P_A|)^2 d\mu(A),
$$
where $d\mu(A)$ is the Haar measure on the complex 
Grassmannian $G(N,2)$. Note $(dd_c \log |P_A|)^2$ is 
well-defined for all $A$ except for an algebraic subset of 
$G(N,2)$ which is zero measure.
Since $P_A$ has exactly two components, it follows from 
the discussion of $N=2$  case that $\int (dd_c \log |P_A|)^2$ 
is given by the intersection number of $P_1^A$ and $P_2^A$, 
which is an integer.
This means $A\mapsto \int (dd_c \log |P_A|)^2$ is an 
integer-valued continuous function. We have now proved 
that the value of 
$I_{\rm intersection}$ is quantized.
}

This concludes our discussion of the topological charges. Our discussion mainly concentrates on the case $N=1$. As we have seen, however, even for $N>1$ case, the overall $U(1)$ part represented by $c_1 = - \frac{1}{2\pi} \Tr \, F$ is still described by the language of the amoeba and tropical geometry. At the same time, we should not think that is the whole story. Out of all the topological charges \eqref{eq:vortex-charge}, \eqref{eq:intersection} and \eqref{eq:instantonnum}, the instanton number
\beq
I_{\rm instanton} ~=~ \int c_2 ~=~ \frac{1}{8\pi^2} \int \left[\Tr \, F \wedge \Tr \, F -\Tr (F\wedge F)\right]
\eeq
vanishes in $U(1)$ theory and appears only in non-Abelian theory. We will discuss this in the next section, but before that let us discuss metric of moduli space for the web of vortices.

\subsection{Metric on Moduli Space}\label{sec:metric}
In this section, we discuss the metric of the moduli space. 
The metric of the moduli space is given by a formula 
similar to the one-dimensional case, 
Eq.\,(\ref{eq:1dim-Kahler1}) and Eq.\,(\ref{eq:1dim-Kahler2}),
\beq
\! \! \! \! \! 
K_{i \bar j} = c \!  \int  \! \left( \frac{\omega^2}{2!} \p_i \bar 
\p_j \log \det \Omega + 2 l^2 \, \omega \wedge i \tr 
\Big[ \bar \p (\Omega \p \Omega^{-1}) 
\bar \p_j (\Omega \p_i \Omega^{-1}) 
- \bar \p (\Omega \p_i \Omega^{-1}) 
\bar \p_j (\Omega \p \Omega^{-1}) \Big] \right)
\label{eq:2dim-Kahler}
\eeq
This is a new result which is valid for arbitrary values of 
$N_{\rm C}$ and $N_{F}$. 

Here we focus on the case $\NC=\NF=1$ in which all the moduli parameters are contained in the Laurent polynomial. 
Let us consider a Newton polytope $\Delta(P)$ associated with a Laurent polynomial $P$. Let $V_{ex}(Q)$ and $V_{in}(Q)$ be sets of external and internal vertices of $\Delta(P)$, respectively. The coefficients of the Laurent polynomial $P$ are identified with the moduli parameters and there exist zero modes corresponding to these moduli parameters. 
\beq
P(u_1,u_2) = \sum_{(n_1,n_2) \in V(Q)} a_{n_1,n_2} u_1^{n_1} u_2^{n_2} .
\label{eq:LP}
\eeq
The coefficients $a_{n_1,n_2},~(n_1,n_2) \in V_{ex}(Q)$ 
determine the positions of the external legs and the size 
of the loops of the vortex web, 
while the coefficients $a_{n_1,n_2},~(n_1,n_2) \in V_{in}(Q)$ 
determine only the size of the loops of the vortex web. 
Since the zero modes associated to the motion of the 
external legs are non-normalizable, we cannot define the 
metric for these zero modes and we must fix the moduli 
parameters $a_{n_1,n_2},~(n_1,n_2) \in V_{ex}(Q)$. 

First let us consider the case where the loop sizes and the 
radii of the torus are much larger than the width of the 
vortex sheets $l \equiv 1/g\sqrt{c}$. 
In this case we can evaluate the leading terms in the 
K\"ahler metric by taking the thin vortex sheet limit 
$l \rightarrow 0$ 
\beq
\lim_{l \rightarrow 0} K_{i \bar j} 
= 2c \int d^4 x \, \p_i \bar \p_j \log |P|.
\label{eq:metric-infinite}
\eeq
From Eq.(\ref{eq:Ronkin}) we obtain the K\"ahler potential 
in the thin vortex sheet limit $l \rightarrow \infty$ as 
\beq
K \approx 8 \pi^2 c R_1 R_2 \int d^2 x 
\left( N_P(x_1,x_2,a,\bar a) - f(x_1,x_2,a) 
- \overline{f(x_1,x_2,a)} \right).
\label{eq:Kahler1}
\eeq
Here $f(x_1,x_2,a)$ is a holomorphic function with respect 
to $a_{n_1,n_2},~(n_1,n_2) \in V_{in}(Q)$ and it should be 
chosen to make the K\"ahler potential finite. 
A possible choice of the function $f(x_1,x_2,a)$ is 
\beq
f(x_1,x_2) &=& \frac{1}{2} N_{\widetilde P}(x_1,x_2), \label{eq:2dim-f}\\
\widetilde P(u_1,u_2) &=& \sum_{(n_1,n_2) \in V_{ex}(Q)} a_{n_1,n_2} u_1^{n_1} u_2^{n_2}.
\label{eq:tilde-P}
\eeq
Therefore the K\"ahler potential can be evaluated 
by integrating the Ronkin functions. 

Next let us consider the case where the loop sizes are much larger than the radius of torus $R \equiv R_1 = R_2 \gg l$. In this case the leading terms in the K\"ahler potential can be evaluated by taking the small radius limit of Eq.\,(\ref{eq:Kahler1}). Since $K \approx {\mathcal O}(R)$ in the limit $R \rightarrow 0$, the leading term in the K\"ahler potential is given by
\beq
K &\approx& 8 \pi^2 c R \lim_{R \rightarrow 0} \int d^2 x R \left( N_P(x_1,x_2,a,\bar a) - N_{\widetilde P}(x_1,x_2) \right) \notag \\
&=& 8 \pi^2 c R \int d^2 x \left( F_P(x_1,x_2) - F_{\widetilde P}(x_1,x_2) \right),
\label{eq:2dim-asymptotic-KP}
\eeq
where $F_{P}$ and $F_{\widetilde P}$ are piece-wise linear functions defined by
\beq
F_P &=& \lim_{R \rightarrow 0} R N_{P}(x_1,x_2) = \underset{(n_1,n_2) \in V(Q)}{\rm max}(n_1 x_1 + n_2 x_2 + r_{n_1,n_2}), \\
F_{\widetilde P} &=& \lim_{R \rightarrow 0} R N_{\widetilde P}(x_1,x_2) = \underset{(n_1,n_2) \in V_{ex}(Q)}{\rm max}(n_1 x_1 + n_2 x_2 + r_{n_1,n_2}),
\label{eq:2dim-Fp}
\eeq
with $r_{n_1,n_2} \equiv R \log |a_{n_1,n_2}|$. 

\begin{figure}[htbp]
\begin{center}
\includegraphics[width=80mm]{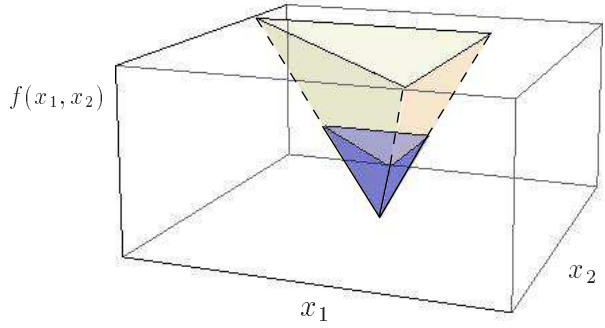} \hs{5}
\caption{The asymptotic K\"ahler potential is proportional to the volume of the tetrahedron surrounded by four planes $f(x_1,x_2) = -x_1-x_2$, $f(x_1,x_2) = x_1$, $f(x_1,x_2) = x_2$, $f(x_1,x_2) = r$.}
\label{fig:tetra}
\end{center}
\end{figure}

Let us consider the simplest example of one-loop 
configuration associated with the Laurent polynomial 
\beq
P(u_1,u_2) = u_1 + u_2 + u_1^{-1} u_2^{-1} + a_{0,0}.
\label{eq:1loop-LP}
\eeq
\begin{figure}[htbp]
\begin{center}
\begin{tabular}{ccc}
\includegraphics[width=50mm]{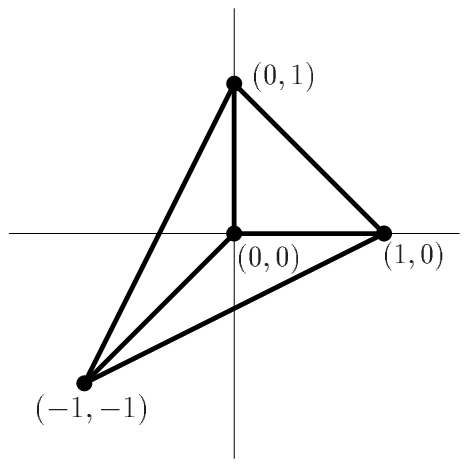} &
\includegraphics[width=50mm]{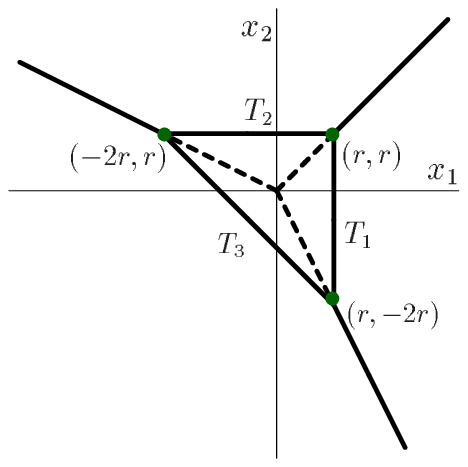} & \\
(a) Newton polytope & (b) web diagram (tropical variety) 
\end{tabular}
\end{center}
\caption{Newton polytope and web diagram for $P=u_1+u_2+u_1^{-1} u_2^{-1} + a_{0,0}$. 
The loop consists of three vortices, 
which can be interpreted as walls 
with tension $T_1 = T_2 = \hat c/R$ and $T_3 = \sqrt{2} \hat c/R$, 
see Eq.\,(\ref{eq:wall-tension})}. 
\label{fig:one-loop}
\end{figure}
In this case the vertices of the Newton polytope are 
\beq
V(Q) &=& \left\{ (-1,-1),\, (0,0),\, (1,0),\, (0,1) \right\} \\
V_{in}(Q) &=& \left\{ (0,0) \right\} \\
V_{ex}(Q) &=& \left\{ (-1,-1),\, (1,0),\, (0,1) \right\} 
\eeq
There exists only one normalizable moduli parameter $a_{0,0}$ for which we can define the metric. This moduli parameter $a_{0,0}$ is related to the size of the loop in Fig.\,\ref{fig:one-loop}-(b), which is proportional to $r \equiv R \log |a_{0,0}|$. The non-normalizable moduli $a_{-1,-1},\,a_{1,0},\,a_{0,1}$ have already been fixed to 1 in Eq.\,(\ref{eq:1loop-LP}). The Laurent polynomial defined in Eq.\,(\ref{eq:tilde-P}) is given by
\beq
\widetilde P(u_1,u_2) = u_1 + u_2 + u_1^{-1} u_2^{-1}. 
\label{eq:1loop-counter}
\eeq
The piece-wise linear functions $F_P(x_1,x_2)$ and $F_{\widetilde P}(x_1,x_2)$ defined in Eq.\,(\ref{eq:2dim-Fp}) are now given by
\beq
F_P(x_1,x_2) = {\rm max}( -x_1 - x_2,\, x_1,\, x_2,\, r), \hs{5}
F_{\widetilde P}(x_1,x_2) = {\rm max}( -x_1 - x_2,\, x_1,\, x_2),
\label{eq:1loop-Fp}
\eeq
By integrating Eq.\,(\ref{eq:2dim-asymptotic-KP}), we obtain the asymptotic K\"ahler potential as the volume of a tetrahedron in Fig.\,\ref{fig:tetra}, which is given by
\beq
K \approx 12 \pi^2 c R \, r^3.
\label{eq:1loop-aymptotic-KP}
\eeq
Differentiating the K\"ahler potential Eq.\,(\ref{eq:1loop-aymptotic-KP}), we obtain the metric of the moduli space and then the effective Lagrangian 
\beq
L_{\rm eff} = 18 \pi^2 c r R \left( \dot r^2 + R^2 \dot \theta^2 \right).
\label{eq:Leff}
\eeq
Here $\theta$ is the phase of the moduli parameter 
$a = e^{r/R + i \theta}$. 
This asymptotic Lagrangian can be interpreted as the 
kinetic energy associated with the motion of the three  
vortices composing the loop. 
Since they can be interpreted as walls 
with tension $T_1 = T_2 = \hat c/R$ and $T_3 = \sqrt{2} \hat c/R$, 
two of three vortices have the mass 
$m_{1,2} = T_{1,2} \times \mbox{length} = 3 \hat c r/R$ 
and the other vortex has the mass 
$m_3 = T_3 \times \mbox{length} = 6 \hat c r /R$. 
If the moduli parameter varies with velocity $\dot r$, the  
vortices move in the $(x_1,x_2)$-plane with velocities 
$v_{1,2} = \dot r$ and $v_3 = \dot r /\sqrt{2}$. 
Then the total kinetic energy associated with the motion 
of the vortices 
is $\sum_{i=1}^3 \frac{m_i}{2} v_i^2 
= \frac{9}{2} \hat c r / R \dot r^2 = 18 \pi^2 c r R \dot r^2$.

\section{Instantons inside Non-Abelian Vortex Webs}\label{sec:instanton}

In this section, we discuss the instanton number $I_{\rm instanton} = \int c_2$ as promised in the previous section. 
We consider the case of $U(2)$ gauge theory with $\NF=2$ scalar fields as the simplest model which admits the BPS configuration with $I_{\rm instanton} \not = 0$. The generalization to the $U(N)$ gauge group should be straightforward.

\subsection{Instanton Number on a Planar Vortex Plane: a Review}
\label{sec:i-number1}
Let us first review the 1/2 BPS vortex moduli space. 
The vacuum of $U(2)$ gauge theory with $\NF=2$ scalar fields 
breaks the color and flavor symmetry $U(2)_{\rm C} \times SU(2)_{\rm F}$ 
into color-flavor locked symmetry $SU(2)_{\rm C+F}$.
The 1/2 BPS single vortex in this theory further breaks 
$SU(2)_{\rm C+F}$ into $U(1)_{\rm C+F}$. 
Therefore there appear 
Nambu-Goldstone modes of the complex projective space 
$\C P^1 \simeq SU(2)_{\rm C+F}/U(1)_{\rm C+F}$ 
localized around the vortex \cite{Hanany:2003hp,Auzzi:2003fs}. 
The orientation moduli $\mathcal M_{\rm orientation} \equiv \C P^1$ 
form a part of the moduli space. 

The moduli matrix for the vortex at $z_1=0$ 
with an orientational moduli parameter $b$ is given by \cite{Eto:2005yh}
\beq
H_0 = \sqrt{c} \ba{cc} 1 & b \\ 0 & z_1 \ea \sim \sqrt{c} \ba{cc} z_1 & 0 \\ 1/b & 1 \ea,
 \label{eq:MM2planar}
\eeq
where $\sim$ represents the $V$-equivalence relation.
These two moduli matrices provide two patches $b$ and $1/b$ 
of $\C P^1$ \cite{Eto:2005yh}. 
For this moduli matrix, the solution of the master equation is given by
\beq
\Omega = \ba{cc} 1 + |b|^2 & b \bar z_1 \\ \bar b z_1 & \frac{\Omega_\ast - |z_1|^2}{1+|b|^2} + |z_1|^2 \ea,
\label{eq:Omega2planar}
\eeq
where $\Omega_\ast$ is a solution of the master equation in 
the case of the Abelian-Higgs model ($\NC=\NF=1$),
such that
\beq
\p_{\bar z_1} (\Omega_\ast \p_{z_1} \Omega_\ast^{-1}) = - \frac{g^2c}{4} ( 1 - |z_1|^2 \Omega_\ast^{-1}).
\label{eq:masterANO}
\eeq
Therefore once the vortex solution $\Omega_\ast$ 
of the Abelian-Higgs model is given, 
one can construct the whole solution of the non-Abelian model. 

Next let us discuss the 1/4 BPS configuration of 
instantons and vortices on $(\C^*)^2$. 
The 1/2 BPS vortex plays the role of a host soliton 
in the 1/4 BPS vortex-instanton configuration. 
It has an internal degree of freedom parametrized by an orientational moduli parameter $b$, which is interpreted as an inhomogeneous coordinate of 
$\mathcal M_{\rm orientation} = \C P^1$ as denoted above.  
The instantons inside the vortex can be constructed as 
lumps in the vortex effective theory which is 
the $\C P^1$ sigma model \cite{Hanany:2004ea,Eto:2004rz}. 
By using the moduli matrix (\ref{eq:MM2planar}) 
for 1/2 BPS single vortex, 
we can construct the moduli matrix for some 1/4 BPS configurations as follows. 
The moduli matrix for the 1/4 BPS vortex-instanton 
configuration is obtained by promoting the orientational 
moduli $b$ to a holomorphic function of the other 
holomorphic coordinate $z_2$ \cite{Eto:2004rz}
\beq
    H_0 = \sqrt{c} \ba{cc} 1 & b(z_2) \\ 0 & z_1 \ea, \hs{10} 
 b(z_2) = a \prod_{i=1}^k (z_2 - z_2^{(i)}), 
 \label{eq:moduli-instanton}
\eeq
where the moduli parameter $z_2^{(i)}$ denotes the position 
of the $i$-th instanton 
and $a$ determines the overall size of the instantons 
on the vortex sheet defined by $z_1=0$. 
This function $b(z_2)$ is regarded as a holomorphic 
map from the vortex sheet to 
$\mathcal M_{\rm orientation} = \C P^1$. 

In order to compute the instanton number, 
we require the information upon the solution of 
the master equation $\Omega$ for the moduli matrix 
Eq.\,(\ref{eq:moduli-instanton}). 
Since the topological charges are
determined only from the boundary condition, 
it can be calculated from the asymptotic behavior 
of the solution $\Omega$ which is 
obtained by the following procedure. 
Solving the BPS equation in the 2+1-dimensional 
effective theory on the vortex worldvolume, 
we obtain the lump solution $b(z_2)$ 
corresponding to the instanton inside the vortex sheet.
Substituting back the solution of the effective theory $b(z_2)$
into the 1/2 BPS vortex solution Eq.\,(\ref{eq:Omega2planar}), 
we find 
\beq
\Omega \equiv \ba{cc} 1 + |b(z_2)|^2 & b(z_2) \bar z_1 \\ 
\bar b(z_2) z_1 & \frac{\Omega_\ast-|z_1|^2}{1+|b(z_2)|^2} + |z_1|^2 \ea. 
\label{eq:eff-Omega}
\eeq
Although this is not an exact solution of the master equation since $b(z_2)$ is not a constant, we can see that this matrix $\Omega$ possesses the correct asymptotic behavior at spatial infinity 
by substituting $\Omega$ into the master equation. 
Therefore the topological charge, which is determined by behavior of the fields at spatial infinity, can be evaluated from this matrix $\Omega$. Inserting $F=-i S^{-1} \bar \p (\Omega \p \Omega^{-1}) S$ into Eq.\,(\ref{eq:total-instanton}), we obtain 
\beq
I = \frac{1}{8 \pi^2} \int \tr \left( F \wedge F \right) = \frac{1}{8\pi^2} \int d d_c \log \Omega_\ast(z_1,\bar z_1) \wedge d d_c \log (1+|b(z_2)|^2) = k.
\eeq
Therefore the instanton charge is measured by the degree of the function $b(z_2)$, namely the degree of the holomorphic map from the vortex sheet $\C^*$ to 
$\mathcal M_{\rm orientation} = \C P^1$.

\subsection{Instanton Number on Non-Abelian Vortex Webs}\label{sec:i-number2}
As we have seen in the example above, the instantons can exist on the planar vortex sheets. 
Now we work out more general 1/4 BPS configurations explicitly. 
The planar host vortex sheet can be extended to
a general web of the vortex sheets characterized by $P(u_1,u_2)$
as discussed in the previous sections. 
We therefore replace the lower-right component $z_1$ 
in the moduli matrix (\ref{eq:moduli-instanton}) by $P(u_1,u_2)$. 
Moreover the holomorphic function $b$ in 
(\ref{eq:moduli-instanton}) is also replaced with 
a function of two holomorphic coordinates $z_1$ and $z_2$. 
The moduli matrix for such 1/4 BPS configuration of 
the instantons and vortex sheets on $(\C^\ast)^2$ thus becomes
\beq
H_0 = \ba{cc} 1 & b(u_1,u_2) \\ 0 & P(u_1,u_2) \ea,
\label{eq:IVMM}
\eeq
where $P(u_1,u_2)$ and $b(u_1,u_2)$ are Laurent 
polynomials 
\beq
P(u_1,u_2) = \sum a_{n_1,n_2} u_1^{n_1} u_2^{n_2}, \hs{10}
b(u_1,u_2) = \sum b_{n_1,n_2} u_1^{n_1} u_2^{n_2}.
\label{eq:LPs}
\eeq
Although Eq.\,(\ref{eq:IVMM}) is not the most general form 
of the moduli matrix for the 1/4 BPS configurations, 
we treat this simple form of the moduli matrix
for essential explanation. 
The coefficients of the Laurent polynomials $a_{n_1,n_2}$ and $b_{n_1,n_2}$ are the moduli parameters which give the location of vortex sheets, and the positions and sizes of the instantons, respectively. 

For the moduli matrix Eq.\,(\ref{eq:IVMM}), 
the instanton charge $I$ in Eq.~(\ref{eq:total-instanton}) 
is computed as follows. 
Since the topological charge should not change 
under continuous deformations, 
we can take a strong coupling limit $g\rightarrow \infty$ 
in which the solution of the master equation $\Omega$ approaches to 
\beq
\Omega ~\rightarrow~ \Omega_0 = \frac{1}{c} H_0 H_0^\dagger = \ba{cc} 1 + |b|^2 & b \overline P \\ P \overline b & |P|^2 \ea.
\eeq 
However a direct calculation does not work 
since the vortex sheets become singular in this limit. 
To avoid the calculation involving the singular vortex sheets,
we perform another deformation of the configuration. 
Similarly to the case of the planar vortex sheet, 
let $\Omega$ be a $2 \times 2$ matrix given by 
\beq
\Omega \equiv \ba{cc} 1 + |b|^2 & b \overline P \\ P \overline b & \frac{\Omega_\ast - |P|^2}{1+|b|^2}+|P|^2 \ea, 
\eeq
where $\Omega_\ast$ is a solution to the following equation
\beq
\bar \p_{\bar z_1} ( \Omega_\ast \p_{z_1} \Omega_\ast^{-1} ) + \bar \p_{\bar z_2} ( \Omega_\ast \p_{z_2} \Omega_\ast^{-1} ) = - \frac{g^2c}{4} (1 - |P|^2 \Omega_\ast^{-1}).
\label{eq:Omega_ast}
\eeq
Although the matrix $\Omega$ is not a solution of the 
master equation unless the holomorphic function 
$b(u_1,u_2)$ is constant everywhere, this matrix has the 
correct topological information about the configuration 
similarly to the planar case.
For this matrix $\Omega$, we can show that
\beq
\int ch_2 
&=& \frac{1}{8\pi^2} \int \tr \left[ \bar \p 
( \Omega \p \Omega^{-1} ) \wedge \bar \p 
( \Omega \p \Omega^{-1} ) \right] \notag \\
&=& \frac{1}{16\pi^2} \int \left( d d_c \log \Omega_\ast \wedge d d_c \log (1 + |b|^2) - \frac{1}{2} d d_c \log \Omega_\ast 
\wedge d d_c \log \Omega_\ast \right). 
\eeq
If we take the strong gauge coupling limit 
$g \rightarrow \infty$ in Eq.\,(\ref{eq:Omega_ast}), 
then $\Omega_\ast$ approaches to $|P|^2$. 
Therefore we obtain 
the instanton charge $I$ in Eq.~(\ref{eq:total-instanton}) 
as  
\beq
I = \frac{1}{8 \pi^2} \int \left( d d_c \log |P| 
\wedge d d_c \log (1 + |b|^2) - d d_c \log |P| 
\wedge d d_c \log |P| \right),
\eeq
where the first term gives the instanton number 
$I_{\rm instanton}$ in Eq.~(\ref{eq:instantonnum}) 
and the second term gives the 
intersection charge $I_{\rm intersection}$ 
in Eq.~(\ref{eq:intersection}) 
which has been 
computed in Sec.\,\ref{sec:topological_charge}. 
The instanton number is rewritten by using the Poincar\'e-Lelong formula
\beq
I_{\rm instanton} = \frac{1}{8 \pi^2} \int_{(\C^\ast)^2} d d_c \log |P| \wedge d d_c \log (1 + |b|^2) = \frac{1}{4\pi} \int_X d d_c \log(1+|b|^2),
\eeq
where $X$ denotes the zero locus of $P$ corresponding to the vortex sheets.
Therefore the instanton number is given by the degree of the map $b|_X:X \rightarrow \C P^1$.

To see a distribution of the topological charge density, we take two limits of the parameters: one is a small instanton limit and the other is a small radius limit $R \rightarrow 0$. 
The small instanton limit is realized by taking the limit $b_{n_1,n_2} \rightarrow \infty$ with fixed ratios $b_{n_1,n_2}/b_{\tilde{n}_1,\tilde{n}_2}$. Then the two form $d d_c \log(1+|b|^2) \rightarrow d d_c \log |b|^2$ has a delta function-like support on the zeros of $b(u_1,u_2)$.  
From this fact we find that 
in the small instanton limit 
the instantons are localized at common 
zeros of $b(u_1,u_2)$ and $P(u_1, u_2)$, and 
that the vortex sheets are located at $P(u_1, u_2)=0$. 
We also find that instantons are localized at the positions 
of lumps from the viewpoint of effective theory on 
the vortex sheets. 

\begin{figure}[htbp]
\begin{center}
\begin{tabular}{ccc}
\includegraphics[width=40mm]{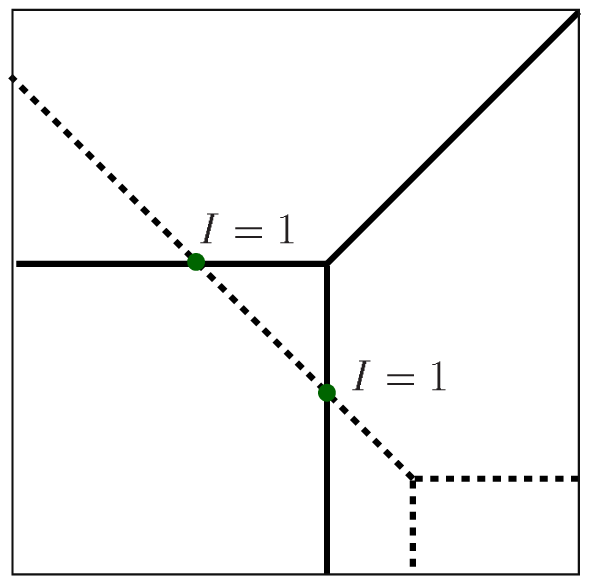} &
\includegraphics[width=40mm]{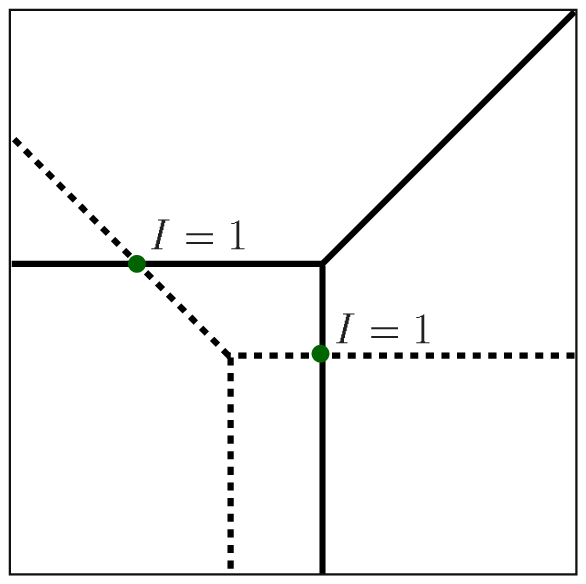} & 
\includegraphics[width=40mm]{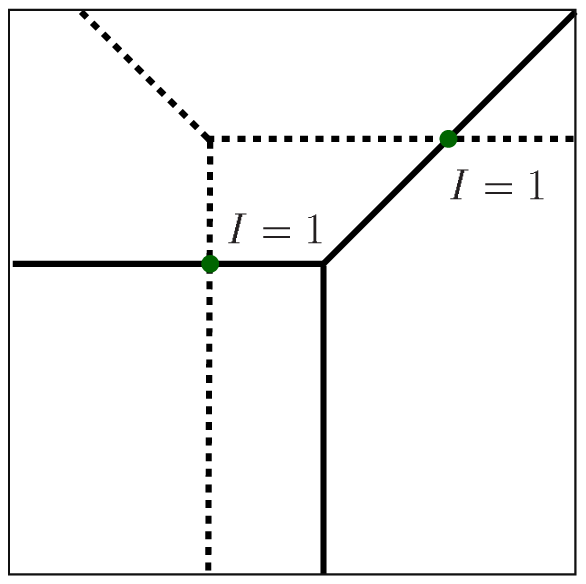} \\
(a) & (b) & (c) 
\end{tabular}
\begin{tabular}{cc}
\includegraphics[width=40mm]{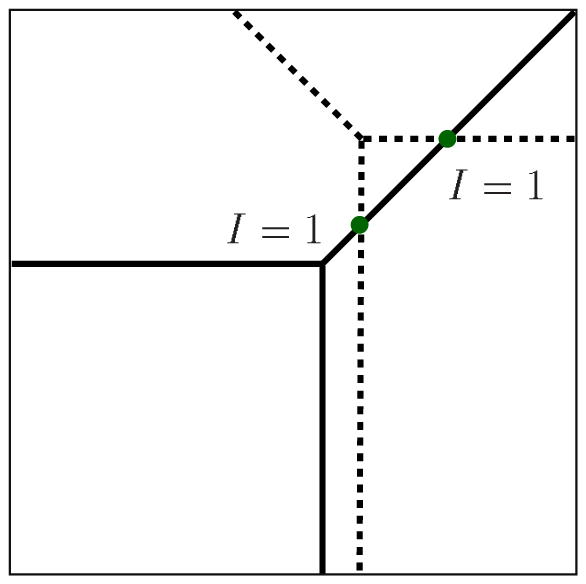} & 
\includegraphics[width=40mm]{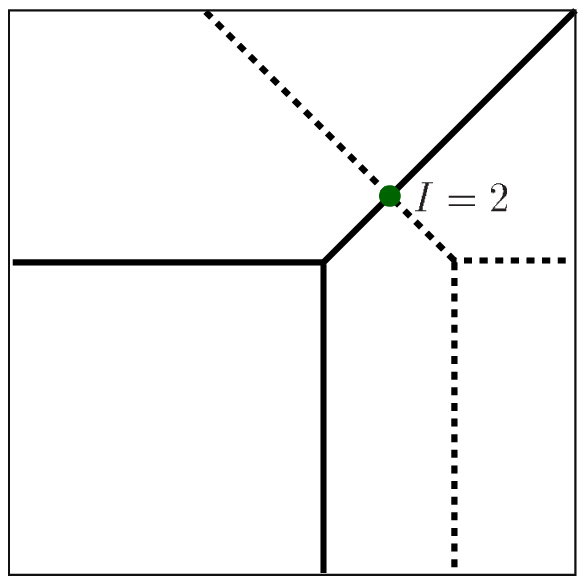} \\
(d) & (e) 
\end{tabular}
\caption{The instanton number density in the small radius limit for $P = u_1 + u_2 + 1$ and $b= b_{1,1} u_1 u_2 + b_{1,0} u_1 + b_{0,0}$. The instanton number density is localized at the intersections of the tropical variety for $P$ (solid line) and the lines on which $\tilde{F}_b(x_1,x_2)$ is not differentiable (dashed lines).}
\label{fig:I-small-radius}
\end{center}
\end{figure}

Next, let us consider the small radius limit $R \rightarrow 0$. 
In this limit, the function $\log(1+|b|^2)$ becomes
\beq
\frac{R}{2} \log (1 + |b|^2) \rightarrow \tilde{F}_b(x_1,x_2) \equiv \underset{(n_1,n_2)}{{\rm max}} (n_1 x_1 + n_2 x_2 + s_{n_1,n_2}), 
\label{eq:piecewise-linear-b}
\eeq
where $s_{0,0} = \frac{R}{2} \log (1+|b_{0,0}|^2)$ and $s_{n_1,n_2} = R \log |b_{n_1,n_2}|$ for $(n_1,n_2)\not=0$ are fixed in the limit. 
Then the instanton number takes the form
\beq
I_{\rm instanton} &=& \frac{1}{8 \pi^2} \int_{(\C^\ast)^2} d d_c \log |P| \wedge d d_c \log (1 + |b|^2) \notag \\
&=& \int_{\R^2} d^2 x \, \epsilon_{ij} \epsilon_{kl} \frac{\p}{\p x_i} \frac{\p}{\p x_k} F_{P}(x_1,x_2) \frac{\p}{\p x_j} \frac{\p}{\p x_l} \tilde{F}_{b}(x_1,x_2).
\label{eq:I-small-radius}
\eeq
Therefore the instanton number density is localized at the intersection of the tropical variety of the Laurent polynomial $P$ and the lines on which the piece-wise linear function $\tilde{F}_b(x_1,x_2)$ is not differentiable. When the instanton number density is localized at the intersection of the lines $n_1 x_1 + n_2 x_2 + r = 0$ and $\tilde{n}_1 x_1 + \tilde{n}_2 x_2 + s = 0$, the instanton number is given by
\beq
I_{\rm instanton} &=& \int d^2 x \, \epsilon_{ij} \epsilon_{kl} \, n_i n_k \delta(n_1 x_1 + n_2 x_2 + r) \, \tilde{n}_j \tilde{n}_l \delta(\tilde{n}_1 x_1 + \tilde{n}_2 x_2 + s) \notag \\
&=& \left| \det \ba{cc} n_1 & n_2 \\ \tilde{n}_1 & \tilde{n}_2 \ea \right| = \left| n_1 \tilde{n}_2-\tilde{n}_1 n_2 \right|.
\eeq
Fig.\,\ref{fig:I-small-radius} shows examples 
of the instanton number density on $\R^2$ 
for $P = u_1 + u_2 + 1$ and 
$b = b_{1,1} u_1 u_2 + b_{1,0} u_1 + b_{0,0}$. 
Varying the moduli parameters,
the instantons localized at the intersection 
move along the tropical variety of $P$. 
For each intersection in Fig.\,\ref{fig:I-small-radius}(a)-(d)
the instanton number is $I=1$. 
In Fig.\,\ref{fig:I-small-radius}(e) 
$I=2$ instanton, which can be interpreted as the coincident 
instantons, is localized at the intersection of the lines 
$x_1-x_2=0$ and $x_1+x_2+R\log|b_{1,1}|-\frac{R}{2}\log(1+|b_{0,0}|^2)=0$. 

So far we have treated the specific configuration of the 
non-Abelian vortex sheets 
and instantons on $(\C^\ast)^2$ and the 
specific gauge group $U(2)$. 
For more general cases, the computation of the instanton number
seems to be complicated and difficult, but the essence should be similar to the above calculations. 

\section{Conclusion and Discussion}\label{sec:discussion}

In this paper, we have investigated generic intersections 
(or webs) of vortices with instantons inside, which is a 
1/4 BPS state in the Higgs phase of (the bosonic part of ) 
five-dimensional $\mathcal{N}=1$ supersymmetric $U(\NC)$ 
gauge theory on $\mathbb{R}_t\times (\C^{\ast})^2 \sim 
\mathbb{R}^{2,1}\times T^2$ with $\NF=\NC$ Higgs scalars 
in the fundamental representation. 
We have found that this vortex-instanton system can be beautifully and naturally understood in the mathematical framework of the amoeba and tropical geometry, and have proposed a dictionary relating the solitons and gauge theory to the amoeba and tropical geometry (summarized in Table \ref{tbl:dictionary}). 

\begin{table}[htbp]
\begin{center}
\caption{Dictionary relating soliton/gauge theory to amoeba/tropical geometry.}
\begin{tabular}{|c|c|}
\hline
soliton/gauge theory & amoeba/tropical geometry \\
\hline

moduli matrix $H_0(z_1,z_2)$ & Newton Polynomial $P(u_1,u_2)$ \\

projection of vortex sheet & amoeba $\mathcal{A}_P$ \\

$R\to 0$ 
& tropical limit \\

position of step-wise kinks & tropical variety\\

Wilson loop $\Tr \Sigma_i$ & derivative of Ronkin function: $\p_i N_P$, 
(\ref{eq:Ronkin})\\

intersection charge $I_{\rm intersection}$& (total mass of) complex Monge-Amp\`ere measure \\

vortex charge density $\mathcal{V}$& Laplacian of Ronkin function: 
(\ref{eq:vortex-Ronkin}) 
\\
\hline 
\end{tabular}
\label{tbl:dictionary}
\end{center}
\end{table}

In this discussion, the moduli matrix formalism 
has played crucial roles. The solutions to 1/4 BPS equations are parametrized by a holomorphic function (Laurent polynomial) $H_0(z_1,z_2)$ of two complex parameters $z_1,z_2$ of $(\C^\ast)^2$. This Laurent polynomial can also be considered as a Newton polynomial of some convex polytope $\Delta$, namely the grid diagram. In the strong gauge coupling limit, the position of vortices is exactly given by  the zero of $H_0$, while the projection of the shape of vortex sheet is the amoeba of $\Delta$. Moreover, we can relate Wilson loops in $T^2$, or the zero modes of gauge fields in Kaluza-Klein decomposition, to the derivatives of the Ronkin function $N_{H_0}(x_1,x_2)$, which is a convex function and is defined from Newton polynomial $H_0$.

The relation with the tropical limit and tropical geometry has also been discussed. In the discussion of solitons, it is natural to consider dimensional reduction of the theory in order to obtain BPS solitons in lower-dimensional field theories. This limit is known in the mathematical literature as the tropical limit. In this limit, the shape of amoeba degenerates into a tropical variety, which is nothing but the so-called $(p,q)$-web of the grid diagram. We have shown that tropical geometry 
provides simple and elegant method to understand not only the dimensionally reduced theory but also the original vortex-instanton system on $(\C^\ast)^2$.

We have also discussed the topological charges, which are 
divided into three types.  
They are the vortex 
charge $V$, the intersection charge 
$I_{\rm intersection}$ 
(negative contribution of the instanton charge) and the 
instanton number $I_{\rm instanton}$. 
First, 
the vortex charge is uniformly distributed along vortex  
sheet det$H_0=0$, 
and its density is given by the Laplacian of the Ronkin 
function in the tropical limit. 
Its total charge is the area of the vortex sheet multiplied 
by $2\pi c$. 
Second, in the strong gauge coupling limit the intersection 
charge density is given by the complex Monge-Amp\`ere 
measure of a plurisubharmonic function $\log |H_0(z_1,z_2)|$, 
and total intersection charge is given by the area of the grid 
diagram $\Delta$ with a suitable regularization. 
Third, the instanton number $I_{\rm instanton}$ 
appears only in the non-Abelian 
case, and we have discussed the case of $\NF=\NC=2$ as an 
example. 
Our discussion simplifies in two limits. 
In the small instanton limit, instantons are localized at 
intersections of $P(z_1,z_2)$ and another Laurent 
polynomial $b(z_1,z_2)$, which parametrizes 
an orientational moduli ${\C}P^1$. 
In the small radius limit, the instanton number 
density ${\cal I}_{\rm instanton}$ is 
localized at the intersection of the tropical varieties corresponding to $P$ and $b$.
We have also obtained the general form of the K\"ahler 
potential and the asymptotic metric of the moduli space of 
a vortex loop as a byproduct of the discussions above. 
In the tropical limit, the K\"ahler potential is given by 
the volume of a convex polytope, and the effective 
Lagrangian can be interpreted as the kinetic energy 
associated with the motion of vortices 
composing the loop.

In our discussion, we mainly focused on the case of 
Abelian-Higgs model ($\NF=\NC=1$), but as far as 
the overall $U(1)$ part 
is concerned the story is exactly the same in non-Abelian $U(\NC)$ case. We also obtained new results by going to non-Abelian gauge group, such as the instanton number. This seems to suggest a non-Abelian generalization of the amoeba and tropical geometry.

\bigskip

Despite such impressive success, there are still many points which need further exploration.

First, in this paper, we have found one-to-one correspondence between 
the amoeba/tropical geometry and solitons in {\it Abelian} gauge theory. 
However the configuration of instantons inside 
non-Abelian vortex-webs in non-Abelian gauge theory 
discussed in Sec.~\ref{sec:instanton} 
does not correspond to the amoeba and tropical geometry so far. 
This configuration suggests non-Abelian generalization of amoeba 
and tropical geometry. 
Furthermore the non-Abelian vortices have been recently 
extended to the case of gauge group $G = U(1) \times G'$ 
with $G'$ {\it arbitrary} simple group \cite{Eto:2008yi}. 
This is the case of complex one dimension. 
It should be extendible to the case of complex two dimensions 
such as $(\C^*)^2$. 
That may suggest further generalization of the amoeba and 
tropical geometry associated to arbitrary group, which contains the usual one as a special case of $U(1)$ gauge group.%
\footnote{In fact, the motto of tropical geometry is to extend usual algebraic geometry by replacing commutative ring with a commutative semiring. Another generalization is to replace a commutative ring with a non-commutative ring, which is non-commutative geometry. Perhaps our discussion of non-Abelian vortices suggests further generalization by combining above two, which should be called ``non-commutative/non-Abelian tropical geometry''.}

Second, let us pursue a possibility to generalize the space where 
the amoeba lives. 
In this paper we have considered the amoeba on $(\C^*)^2$. 
This is because the maximal space-time dimension of 
supersymmetric gauge theory with eight supercharges 
is $d=5+1$. 
Since space-dimension $5$ is odd we have studied four dimensional case 
$(\C^*)^2$. 
If we abandon implementing supersymmetry, we can extend the 
bosonic Lagrangian (\ref{eq:Lagrangian}) to space 
dimensions higher than $5$, 
and study higher co-dimensional
composite solitons of the vortices and instantons 
extending to various directions. 
In fact the generalized vortex equations 
on arbitrary K\"ahler manifold of arbitrary dimensions
were obtained in (bosonic) Yang-Mills-Higgs theory 
\cite{MundetiRiera:1999fd}. 
Therefore we expect those equations give further 
correspondence of the amoeba and gauge theory 
on $(\C^*)^n$ or on general K\"ahler manifolds.

The third topic is the relation with dimer model \cite{KenyonOkounkov,KOS}. Dimers do not appear directly in our discussion, but it is known that dimer model is intimately connected with the amoeba and tropical geometry. 
For example, the Ronkin function as defined in \eqref{eq:Ronkin} coincides with the thermodynamic limit of partition function of a dimer model. Moreover, a spectral curve of the dimer model is known \cite{KenyonOkounkov} to parametrize Harnack curve, whose amoeba has good properties. 
The dimers also appear in discussion of the brane tilings \cite{tiling,tilingreview} and four-dimensional $\mathcal{N}=1$ superconformal quiver gauge theories. The brane tilings now have an interpretation as configuration of D5-branes and NS5-brane \cite{tilingasbrane}, whose brane configuration is shown in Table \ref{tbl:D5NS5},
\begin{table}[htpb]
\begin{center}
\caption{The five-brane configuration described by brane tilings. In the weak string coupling limit, the surface $\Sigma$ becomes a zero locus of a Newton polynomial corresponding to the toric diagram. This setup is analogous to our vortex-instanton systems, although details are different.}
\begin{tabular}{c|cccc|cccc|cc}
\hline
\hline
&0&1&2&3&4&5&6&7&8&9 \\
\hline
D5&$\circ$ &$\circ$ &$\circ$ &$\circ$ & & $\circ$ & &$\circ$& & \\
NS5&$\circ$ &$\circ$ &$\circ$ &$\circ$ &\multicolumn{4}{c}{$\Sigma$ (2-dim surface)} & \multicolumn{1}{|c}{} \\
\hline
\end{tabular}
\label{tbl:D5NS5}
\end{center}
\end{table} 
wherein $\Sigma$ is a two-dimensional surface in 4567-directions. In the weak gauge coupling limit, this $\Sigma$ is the zero locus of a Newton polynomial with respect to two complex variables in $(\mathbb{C}^\ast)^2$. Hence the mathematical structure of the NS5-brane is exactly the same as that of the vortex sheets we have considered in this paper. The toric diagram and the grid diagram are identical, and the meaning of the tropical limit and tropical variety also coincides. Of course, we should keep in mind that there exists important differences between the soliton systems in this paper and the brane tilings. First, in this paper we have assumed systems with eight supercharges, but in brane tiling, we have only four supercharges. Related to this fact is that we do not have analogue of D5-brane in the soliton side, and have the instanton charge instead. Still, we might obtain something new from this analogy. For example, in the discussion of brane tilings, the projection of $(\C^\ast)^2$ on to $T^2$ directions, which is called alga or coamoeba, plays crucial roles \cite{Feng:2005gw,UY}. It would be interesting to see whether coamoeba has any significance in our setup. 
Perhaps we can understand these points better if we can 
find a D-brane realization of the vortex-instanton system. 
 For this direction, the work on D-brane configuration of 
vortices on cylinder \cite{Eto:2006mz,Eto:2007aw}, 
its T-dual to the D-brane configuration of the domain 
walls\cite{Lambert:1999ix,Eto:2004vy}, and the D-brane 
configurations of the domain wall webs \cite{Eto:2005mx} 
should be a useful guideline. 

The amoeba and tropical geometry also appear in the computation of the topological string amplitude and the instanton counting in 4d and 5d supersymmetric gauge theories. Their perturbative dynamics are ruled by asymptotic behavior of the (plane) partitions where the amoeba and the Ronkin function appear. It is also pointed out that the K\"ahler structure behind the theories is closely related with the volume of the convex cone of the Ronkin function 
similarly to our discussion on the K\"ahler potential.
These relations suggest that the intersecting soliton 
system also admits an interpretation of the microscopic 
partitions (dimers) and that the K\"ahler geometry
of the moduli space of the solitons is determined by a suitable asymptotic limit of the microscopic interpretation.

 Finally, statistical partition functions of the vortices 
on a cylinder were studied by using D-brane configurations 
\cite{Eto:2007aw}. 
There the integration over the moduli space of vortices 
is drastically simplified in the T-dual picture: vortices 
are mapped to domain walls and the integration reduces 
to a problem of rods. 
The limit of parameters employed in \cite{Eto:2007aw} is a 
little different from that in this paper. There the limit 
$g \rightarrow \infty,~R \rightarrow 0$ was taken with 
fixed wall width $d \equiv \frac{2}{g^2 c R}$, while in 
this paper we have taken the strong coupling limit 
$g \rightarrow \infty$ first and then the small radius 
limit $R \rightarrow 0$. 
This is nothing but the non-linear sigma model version of 
the limit in \cite{Eto:2007aw}. 
In the case of $\NF > \NC$, the gauged linear sigma model 
reduces to the non-linear sigma model and the vortex 
solution reduces to the lump solution \cite{Eto:2007yv}. 
The lump solution on the cylinder can be mapped to kinks 
on one-dimensional space. 
The small radius limit $R \rightarrow 0$, which corresponds 
to dequantization (ultradiscretization) limit 
$\lim_{R \rightarrow 0} R \log( e^{A/R} + e^{B/R}) = \mbox{max} (A, B)$, 
enable us to identify the kinks with free particles in 
one-dimensional space as in Sec.\,\ref{sec:cylinder}. 
The partition function of the free particles gives the 
exact volume of the moduli space of the sigma-model lumps, 
namely the partition function for a multi-lump system. 
This fact and the result of \cite{Eto:2007aw} suggest that 
the procedure of the dequantization is powerful enough to 
give the exact partition function of the solitons in 
non-linear sigma model and variant of it makes the 
computation of the partition function very simple even in 
the case of finite gauge couplings. 
This should be extendible to the vortex webs 
discussed in this paper, to obtain a partition function of 
the instanton-vortex system. 
We expect that it will be reduced to 
the Nekrasov's partition function 
in the limit of 
$g\sqrt{c}\to0$
where the vortices disappear while the instantons still remain. 
The integration over the moduli space of the instanton-vortex system 
may provide a systematic method to compute 
the symplectic Gromov-Witten invariant, 
which is a combination of the Donaldson invariant and the Gromov-Witten 
invariant \cite{sympleticGW}.

\section*{Acknowledgements}
We are grateful to Minoru Eto, Takayuki Nagashima and 
Keisuke Ohashi for collaboration in early stages of this project. 
We would like to thank the Yukawa Institute for Theoretical 
Physics at Kyoto University, where this work was initiated 
during the workshop YITP-W-06-16 on ``Foundamental Problems 
and Applications of Quantum Field Theory''. 
M.N. and K.O. are supported in part by Grant-in-Aid for Scientific
Research (No. 20740141 and No.19740120, respectively) from the Ministry
of Education, Culture, Sports, Science and Technology.
T. F. and M.Y. are supported by the JSPS Research Fellowships 
for Young Scientists. 
This work is supported in part by Grant-in-Aid for 
Scientific Research from the Ministry of Education, 
Culture, Sports, Science and Technology, Japan No.17540237
and No.18204024 (N.S.). 
M.Y. would like to thank Yosuke Imamura for 
discussions, 
Alexander Rashkovskii for kind correspondence, and 
Yukawa Institute for Theoretical Physics for
hospitality during the final stages of this work.


\end{document}